\documentclass[%
reprint,
superscriptaddress,
amsmath,amssymb,aps,prx]{revtex4-1}

\usepackage{graphicx}
\usepackage{dcolumn}
\usepackage{bm}
\usepackage{color}

\begin{document}

\preprint{APS/123-QED}

\title{Helicoidal excitonic phase in an electron-hole double layer system}

\author{Ke Chen}
\affiliation{International Center for Quantum Materials, School of Physics, Peking University, Beijing 100871, China}
\affiliation{Collaborative Innovation Center of Quantum Matter, Beijing 100871, China}%
\author{Ryuichi Shindou}%
\email{rshindou@pku.edu.cn}
\affiliation{International Center for Quantum Materials, School of Physics, Peking University, Beijing 100871, China}%
\affiliation{Collaborative Innovation Center of Quantum Matter, Beijing 100871, China}%

\date{\today}

\begin{abstract}
We propose helicoidal excitonic phase in a Coulomb-coupled two-dimensional 
electron-hole double layer (EHDL) system with relativistic spin-orbit interaction. Previously, 
it was demonstrated that layered InAs/AlSb/GaInSb heterostructure is an ideal experimental platform 
for searching excitonic condensate phases, while its electron layer has non-negligible Rashba interaction. 
We clarify that due to the Rashba term, the spin-triplet (spin-1) exciton field in the EHDL system 
forms a helicoidal structure and the helicoid plane can be controlled by an in-plane Zeeman field. 
We show that due to small but finite Dirac term in the heavy hole layer the helicoidal structure 
of the excitonic field under the in-plane field results in a helicoidal {\it magnetic} order in the 
electron layer. Based on linearization analyses, we further calculate momentum-energy 
dispersions of low-energy Goldstone modes in the helicoidal excitonic phase. We discuss 
possible experimental probes of the excitonic phase in the EHDL system.   
\end{abstract}

\pacs{4444}

\maketitle

\section{introduction}
One of the fundamental challenges in condensed matter physics is an experimental realization of 
excitonic condensation and excitonic insulator at the 
equilibrium~\cite{mott61,knox63,keldysh65,jerome67,halperin68,wakisaka09,kogar17,werdehausen18}.  
Early experiments report significant electron-hole Coulomb drag phenomena in bilayer quantum well structure~\cite{lozovik75,comte82,datta85,xia92,zhu95,littlewood96,naveh96} made 
out of semiconductors such as GaAs/AlGaAs~\cite{croxall08,seamons09,yang11} or Si~\cite{sivan92}. 
Recently, a strained layer InAs/AlSb/GaInSb heterostructure~\cite{naveh96,wu19} provides 
an ideal platform of Coulomb coupled electron-hole double layer (EHDL) system, that 
has a lot of advantages over the others. Thereby, a dual gate device enables a continuous change 
of both chemical potential and charge state energy. Large relativistic spin-orbit interaction (SOI) 
realizes a nearly isotropic Fermi contour of the heavy hole band as well as the electron band. 
In fact, a recent experiment reports a transport signature of the excitonic coupling 
in the charge neutrality regime of the strained layer heterostructure~\cite{wu19}. 

The excitonic pairing in Coulomb coupled EHDL system is either of the spin triplet (spin-1) nature or 
of the spin-singlet (spin-0) nature~\cite{zhu95,zhu96}. An energy degeneracy between these two will be 
lifted by the large SOI in the electron layer of the EHDL system. Besides, due to the SOI, 
the spin triplet vector (spin-1 vector) could exhibit a non-trivial spatial texture, 
that breaks the translational symmetry in the two-dimensional 
plane. The spatial texture of the excitonic pairing field in EHDL system is generally free from charged and/or 
magnetic impurities in each layer, while it could be pinned by a spatial variation of the dielectric constant 
in the intermediate separation layer. To our best knowledge, 
it is entirely an open question at this moment how the spin-1 exciton  
forms spatial textures in the presence of the SOI, how the textures could be coupled with external 
magnetic (Zeeman) field, what kind of low-energy collective excitations would emerge from the 
translational symmetry breaking, and how the texture could be experimentally detected. 

In this paper, we identify the spatial textures of the spin-1 exciton condensate in the 
presence of the SOI, clarify the natures of the low-energy collective 
modes in the excitonic phase, and propose possible experimental probes for detecting the 
spatial texture of the spin-1 exciton condensate. First, we derive a $\phi^4$ effective action 
for the spin-triplet (spin-1) excitonic pairing field and determine a form of the couplings among the exciton, 
the SOI and the magnetic Zeeman field. We then propose that helicoidal structures of the spin-1 exciton   
minimize the classical action. Based on the effective action and the classical configuration, we 
derive linearized EOMs for fluctuations of the spin triplet exciton field around the classical configuration. From 
the EOMs, we calculate the low-energy collective excitations in the helicoidal excitonic condensate. 
Using these theoretical knowledges, we propose possible experimental probes for detecting 
the helicoidal excitonic texture in the EHDL system.  

\section{model and effective action}
We begin with a non-interacting Hamiltonian for the two-dimensional EHDL system~\cite{pikulin14};
\begin{align}
H_0 - \mu N &\equiv \int d{\bm x} \!\  {\bm a}^{\dagger}({\bm x}) \bigg[ \Big(-\frac{\hbar^2 \nabla^2}{2m_e} 
- E_{g} - \mu \Big) {\bm \sigma}_0 \nonumber \\ 
&\ \ \  + \xi_e \big(-i\partial_{y} {\bm \sigma}_x 
+ i \partial_x {\bm \sigma}_y\Big) + H {\bm \sigma}_{H} \bigg] {\bm a}({\bm x}) \nonumber \\
& \hspace{-0.2cm} + \int d{\bm x} \!\  {\bm b}^{\dagger}({\bm x}) \bigg[ \Big(\frac{\hbar^2 \nabla^2}{2m_h} 
+ E_{g} - \mu \Big) {\bm \sigma}_0 \nonumber \\ 
&\   + \Delta_h \big(-i\partial_{x} {\bm \sigma}_x 
- i \partial_y {\bm \sigma}_y\Big) + H {\bm \sigma}_{H} \bigg] {\bm b}({\bm x}), \label{hami1}
\end{align}  
with $H=x,y$, ${\bm x} \equiv (x,y)$, ${\bm \sigma}_0$, and ${\bm \sigma}_{i}$ ($i=x,y,z$) 
are two by two unit 
and Pauli matrices. ${\bm a}^{\dagger}({\bm x}) \equiv (a^{\dagger}_{\uparrow}({\bm x}),
a^{\dagger}_{\downarrow}({\bm x}))$ and ${\bm b}^{\dagger}({\bm x}) \equiv 
(b^{\dagger}_{\uparrow}({\bm x}), b^{\dagger}_{\downarrow}({\bm x}))$ denote 
the creation and annihilation operators of electron and hole bands with respective mass 
$m_e \!\ (>0)$ and $m_h \!\ (>0)$. The chemical potential $\mu$ as well as the charge 
state energy $E_g$ (band inversion parameter) can be separately tuned by the dual gate device. 
The SOI in the electron band with $S_z=\pm 1/2$ doublet takes a form of the Rashba term with $\xi_e$ 
and the SOI in the heavy hole band with $J_z=\pm 3/2$ doublet takes a form of the Dirac term with 
$\Delta_h$. In the InAs/AlSb/GaInSb heterostructure, the Rashba term in the electron layer 
is much larger than that in the heavy hole band; $\xi_e/(E_0 d_0)\sim -0.1$, $\Delta_h/(E_0 d_0) \sim 0.001$, 
Here $E_0$ and $d_0$ (thickness of the intermediate separation layer)  define a length scale and energy 
scale of the EHDL system; $E_0=(m^{-1}_e+m^{-1}_h)\hbar^2/2d^2_0=e^2/(4\pi \epsilon \epsilon_0 d_0)$. Typically, 
$E_0/k_{\rm B}= 100 $K and $d_0=10$ nm~\cite{pikulin14,wu19}. For simplicity, 
we take $\Delta_h=0$, unless dictated otherwise (e.g. Section.~V). To study how the spatial texture of the 
spin-1 exciton condenate can be changed by an in-plane magnetic field, we include the magnetic Zeeman 
field $H$ along the in-plane direction ($H=x,y$). 

The electron and hole layers are coupled with 
each other through the long-range Coulomb interaction. For simplicity, we employ the short-ranged 
repulsive interaction; 
\begin{align}
H_{\rm int} = g \int d{\bm x} \!\ a^{\dagger}_{\sigma}({\bm x}) a_{\sigma}({\bm x}) 
\!\ b^{\dagger}_{\sigma^{\prime}}({\bm x})b_{\sigma^{\prime}}({\bm x}). \label{hami2}
\end{align}    
with $g>0$. The spin-rotational symmetry in Eq.~(\ref{hami2}) allows to decompose the interaction into 
the spin singlet and triplet parts. In terms of the respective excitonic pairing fields 
$\hat{\cal O}_{\mu}({\bm x}) \equiv {\bm a}^{\dagger}({\bm x}) 
{\bm \sigma}_{\mu} {\bm b}({\bm x})$ ($\mu=0,x,y,z$), the interaction part takes a form of 
\begin{align}
H_{\rm int} = -\frac{g}{2} \int d{\bm x} \!\ \sum_{\mu=0,x,y,z} 
\hat{\cal O}^{\dagger}_{\mu}({\bm x}) \hat{\cal O}_{\mu} ({\bm x}). \label{hami3}
\end{align}

Using the Stratonovich-Hubbard transformation followed by a standard procedure (appendix A), 
we derive a partition function $Z$ and its action $S$, that is a functional of the spin-triplet (spin-1) 
exciton pairing fields (`$g$-vector'); $\phi_{\mu}({\bm x}) \equiv \frac{g}{2} \langle \hat{\cal O}_{\mu}({\bm x}) \rangle$ 
 $(\mu=x,y,z)$,
\begin{align}
Z &= \int {\cal D}{\bm \phi}^{\dagger} {\cal D}{\bm \phi} \!\ 
\exp \Big[- S[{\bm \phi}^{\dagger},{\bm \phi}]\Big] \nonumber \\
S &= \int^{\beta}_{0} d\tau 
\int d{\bm x} \bigg\{ - \eta {\bm \phi}^{\dagger} \partial_{\tau} {\bm \phi}   
- \Big(\alpha - \frac{2}{g}\Big) |{\bm \phi}|^2 + \lambda |{\bm \nabla} {\bm \phi}|^2 \nonumber \\
& \ \ - \gamma \Big (\big(|{\bm \phi}|^2 \big)^2  + 4 \big(|{\bm \phi}^{\prime}|^2 
|{\bm \phi}^{\prime\prime}|^2  - ({\bm \phi}^{\prime}\cdot {\bm \phi}^{\prime\prime})^2 \big) \Big) 
\nonumber \\  
& \ \ - D \Big( {\bm e}_y \cdot \big({\bm \phi}^{\prime}\times \partial_x {\bm \phi}^{\prime}\big) 
- {\bm e}_x \cdot \big({\bm \phi}^{\prime}\times \partial_y {\bm \phi}^{\prime}\big) \Big) \nonumber \\
& \ \ - D \Big( {\bm e}_y \cdot \big({\bm \phi}^{\prime\prime}\times \partial_x {\bm \phi}^{\prime\prime}\big) 
- {\bm e}_x \cdot \big({\bm \phi}^{\prime\prime}\times \partial_y {\bm \phi}^{\prime\prime}\big) \Big) \nonumber \\
& \ \ - 2h {\bm e}_{H} \cdot \big({\bm \phi}^{\prime}\times {\bm \phi}^{\prime\prime}\big) \bigg\} 
+ {\cal O}(\xi^2_e, H^2, \xi_e H),    \label{action}
\end{align}
with $H=x,y$. ${\bm e}_{x}$ and ${\bm e}_y$ are the unit vectors along $x$ and $y$ axis. 
$\alpha$ and $\lambda$ are positive, and $\eta$ and $\gamma$ are negative (appendix A). 
Since $D$ and $h$ are proportional to $\xi_e$ and $H$ respectively, we can always assume 
$D>0$ and $h>0$ without loss of generality.  
Note that the spin-1 exciton field ${\bm \phi}({\bm x})$ has both real and imaginary parts, 
${\bm \phi}^{\prime}({\bm x})$ and ${\bm \phi}^{\prime\prime}({\bm x})$;  
$\phi_{\nu}({\bm x})\equiv \phi^{\prime}_{\nu}({\bm x})+i \phi^{\prime\prime}_{\nu}({\bm x})$, 
$\phi^{\dagger}_{\nu}({\bm x})\equiv 
\phi^{\prime}_{\nu}({\bm x})-i \phi^{\prime\prime}_{\nu}({\bm x})$ ($\mu=x,y,z$). 
Note also that using a Talyor expansion, we took into account the lowest order in $\xi_e$ and $H$. 
Within the lowest order, the SOI favors helicoid orders of both the real and imaginary parts of the 
$g$-vector, where a rotational plane of the $g$-vector is parallel to a propagating direction within the 
$xy$ plane. The Zeeman field is linearly coupled with a vector chirality formed by 
the real and imaginary parts of the $g$-vector. 

The quadratic part of the action in Eq.~(\ref{action}) gives 
energy dispersions of the spin-1 exciton bands as a function of momentum 
${\bm k}$. The bands are triply degenerate at the zero momentum  
point at the zero magnetic field ($h=0$), while the degeneracy is lifted at non-zero momentum 
due to the SOI term~\cite{yu14}. In the zero field, the lowest spin triplet exciton 
band has a `wine-bottle' minimum at a line (ring) of $|{\bm k}|=K \equiv D/2\lambda$. 
The energy at the minimum is $-\alpha+2/g-D^2/4\lambda$. 
When the energy minimum decreases on lowering temperature or on decreasing the 
charge state energy $E_g$, the minimum eventually touches the zero energy. Thereby, the system picks 
up one ${\bm k}$ from the line of $|{\bm k}|=K$, and undergoes Bose-Einstein condensation 
(BEC)~\cite{pethick01} of the spin-1 exciton band. 
The resulting phase is what we call in this paper {\it helicoidal excitonic condensate phase}. In the next 
section, we assume that $\alpha-2/g+D^2/4\lambda > 0$ and describe this helicoidal excitonic phase in 
details and explain especially how the helicoidal excitonic condensate can be controlled by the 
in-plane Zeeman field ($h=x,y$).

\section{helicoidal excitonic condensate} 
The classical action at the zero magnetic field is maximally minimized by a helicoidal structure of the triplet 
pairing field;
\begin{align}
&{\bm \phi}_{c}({\bm x}) = 
\rho e^{i\theta} \big\{\hat{\bm k} \cos({\bm k}{\bm x}) - \hat{e}_z \sin({\bm k}{\bm x}) \big\}, \label{hs-0-a}  
\end{align}   
with 
\begin{align}
{\bm k} &\equiv K \hat{\bm k} = \frac{D}{2\lambda} \big(\cos \omega \!\ \hat{\bm e}_x + 
\sin \omega \!\ \hat{\bm e}_y\big), \label{hs-0-b} \\ 
K & \equiv \frac{D}{2\lambda}, \ 
\rho \equiv \sqrt{\frac{1}{2|\gamma|} \Big(\alpha-\frac{2}{g} + \frac{D^2}{4\lambda}\Big)}. \label{hs-0-c}
\end{align} 
The spatial pitch of the helicoid structure $1/K$ is determined by the SOI energy ($D$) and the phase stiffness 
energy ($\lambda$) in the classical action. The propagation direction of the helicoid structure $\hat{\bm k}$ is 
arbitrary at the zero field, where the system is symmetric under the continuous rotation of spin and coordinate 
around the $z$-axis; the U(1) phase $\omega$ is arbitrary. The momentum ${\bm k}$ and the $z$-axis 
subtend a rotational plane of the $g$-vector, while the real and imaginary parts of the $g$-vector are parallel to 
each other everywhere (Fig.~\ref{fig:2}(a)). The arbitrary U(1) phase $\theta$ represents a relative gauge degree 
of freedom; a difference between the two U(1) gauge degrees of freedom of the electron and hole bands. 

\begin{figure}[t]
	\centering
	\includegraphics[width=0.8\linewidth]{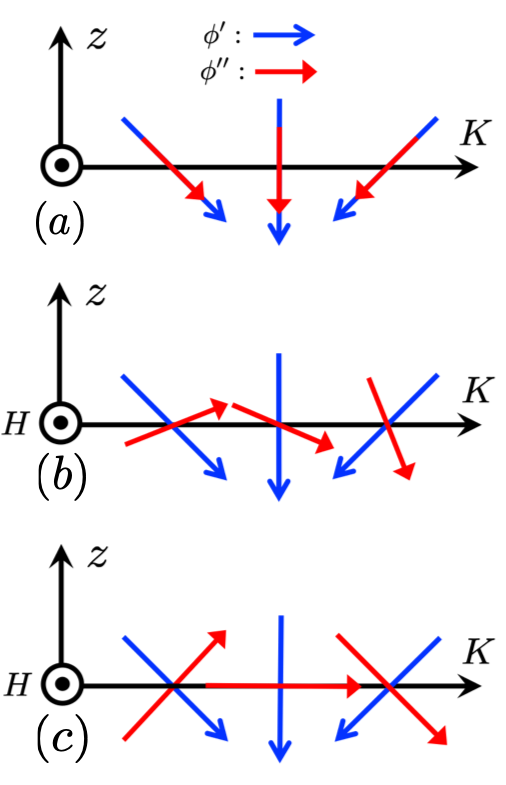}
	\caption{(color online) Helicoidal structure of the spin-triplet (spin-1) exciton field, ${\bm \phi}({\bm x})
 \equiv {\bm \phi}^{\prime}({\bm x}) + i{\bm \phi}^{\prime\prime}({\bm x})$. 
The real/imaginary parts ${\bm \phi}^{\prime}$/${\bm \phi}^{\prime\prime}$ are  
depicted by blue/red arrow respectively. (a) helicoidal structure at the zero field (b) helicoidal structure at $h<h_c$, 
where the in-plane Zeen field is perpendicular to the paper. (c) helicoidal structure at $h>h_c$. The angle between 
the real and imaginary parts is $\pi/2$ and their amplitudes are same everywhere.} 
	\label{fig:1} 
\end{figure}

Under the in-plane Zeeman field (appendix B), the direction of the momentum 
$\hat{\bm k}$ becomes perpendicular to the Zeeman field, 
so that the $g$-vector can rotate around the in-plane Zeeman field (see Fig.~\ref{fig:2}(b,c)). 
The real and imaginary parts of the $g$-vector form 
a finite vector chirality along the field direction. The vector chirality is 
spatially uniform. The amplitude of the vector chirality becomes larger for the larger in-plane Zeeman field. 
Meanwhile, an angle between the real and imaginary part, $\nu$, saturates into $\pi/2$ at a critical field $h_c$ 
defined by   
\begin{align}
h_c \equiv  \alpha-\frac{2}{g} + \frac{D^2}{4\lambda}. \label{satu}
\end{align} 
For an in-plane Zeeman field below the critical field ($h\le h_c$), the helicoid structure of the $g$-vector is given by 
${\bm \phi}_c({\bm x}) = {\bm \phi}^{\prime}_c({\bm x}) + i {\bm \phi}^{\prime\prime}_c({\bm x})$ with 
\begin{align}
&{\bm \phi}^{\prime}_c({\bm x}) = \rho \cos\theta \big(\cos(Ky) \!\ \hat{\bm e}_y - \sin(Ky) \!\ \hat{\bm e}_z\big), \label{hs-1-a} \\ 
&{\bm \phi}^{\prime\prime}_c({\bm x}) = 
\rho \sin\theta \big( \cos(Ky-\nu) \!\ \hat{\bm e}_y - \sin(Ky-\nu) \!\ \hat{\bm e}_z\big). \label{hs-1-b}  
\end{align}
Without loss of generality, we always take the field along the $x$-axis henceforth. The vector chirality 
formed by ${\bm \phi}^{\prime}_c$ and ${\bm \phi}^{\prime\prime}_c$ is spatially uniform, 
and it increases on increasing the field;
\begin{align}
\sin2\theta \sin \nu = \frac{h}{h_c}. \label{hs-1-c}
\end{align} 
A ratio between $|{\bm \phi}^{\prime}_c|$ and $|{\bm \phi}^{\prime\prime}_c|$ is specified by $\theta$, 
and the angle between ${\bm \phi}^{\prime}_c$ and ${\bm \phi}^{\prime\prime}_c$ is specified by $\nu$. 
$\theta$ and $\nu$ form an energy degeneracy line under a fixed vector chirality [Eq.~(\ref{hs-1-c})]. 
When the in-plane field reaches the critical field $h_c$, the angle becomes $\pi/2$ 
and the ratio becomes the unit. Above the critical field ($h\ge h_c$), the angle $\nu$ takes $\pi/2$ and 
the ratio takes one everywhere (Fig.~\ref{fig:2}(c));
\begin{align}
{\bm \phi}_c({\bm x}) 
= e^{iKy} \frac{\rho^{\prime}}{\sqrt{2}} (\hat{\bm e}_y + i\hat{\bm e}_z). \label{hs-2-a}
\end{align}   
with 
\begin{align}
\rho^{\prime} \equiv \sqrt{\frac{h+h_c}{4|\gamma|}}.  \label{hs-2-b}
\end{align} 

\section{low-energy collective modes}
The helicoidal excitonic condensations described in the previous section 
break the spatial translational symmetry, spin rotational symmetry and the relative U(1)
gauge symmetry; the condensate phases are accompanied by gapless Goldstone modes. 
Experimental observation of the collective excitations would serve as future `smoking-gun'  
experiment for the confirmation of the excitonic condensation at the equilibrium 
and therefore, it is important to characterize theoretically energy-momentum dispersion 
of the low-energy collective modes. To this end, we take a functional derivative of the 
effective action (Eq.~(\ref{action})) with respect to ${\bm \phi}$ 
and ${\bm \phi}^{\dagger}$, and derive a coupled non-linear equation of motions (EOMs) for 
the spin-triplet pairing field. The helicoidal structures described in the 
previous section are static solutions of these coupled EOMs. Thus, we consider 
a small fluctuation of the excitonic pairing field around these static solutions, 
${\bm \phi}({\bm x}) = {\bm \phi}_c({\bm x}) 
+ \delta {\bm \phi}({\bm x})$ and ${\bm \phi}^{\dagger}({\bm x}) = {\bm \phi}^{\dagger}_c({\bm x}) 
+ \delta {\bm \phi}^{\dagger}({\bm x})$ and linearize the EOMs with respect to the fluctuation field, 
$\delta {\bm \phi}({\bm x})$ and $\delta {\bm \phi}^{\dagger}({\bm x})$.

\begin{figure}[t]
	\centering
	\includegraphics[width=0.8\linewidth]{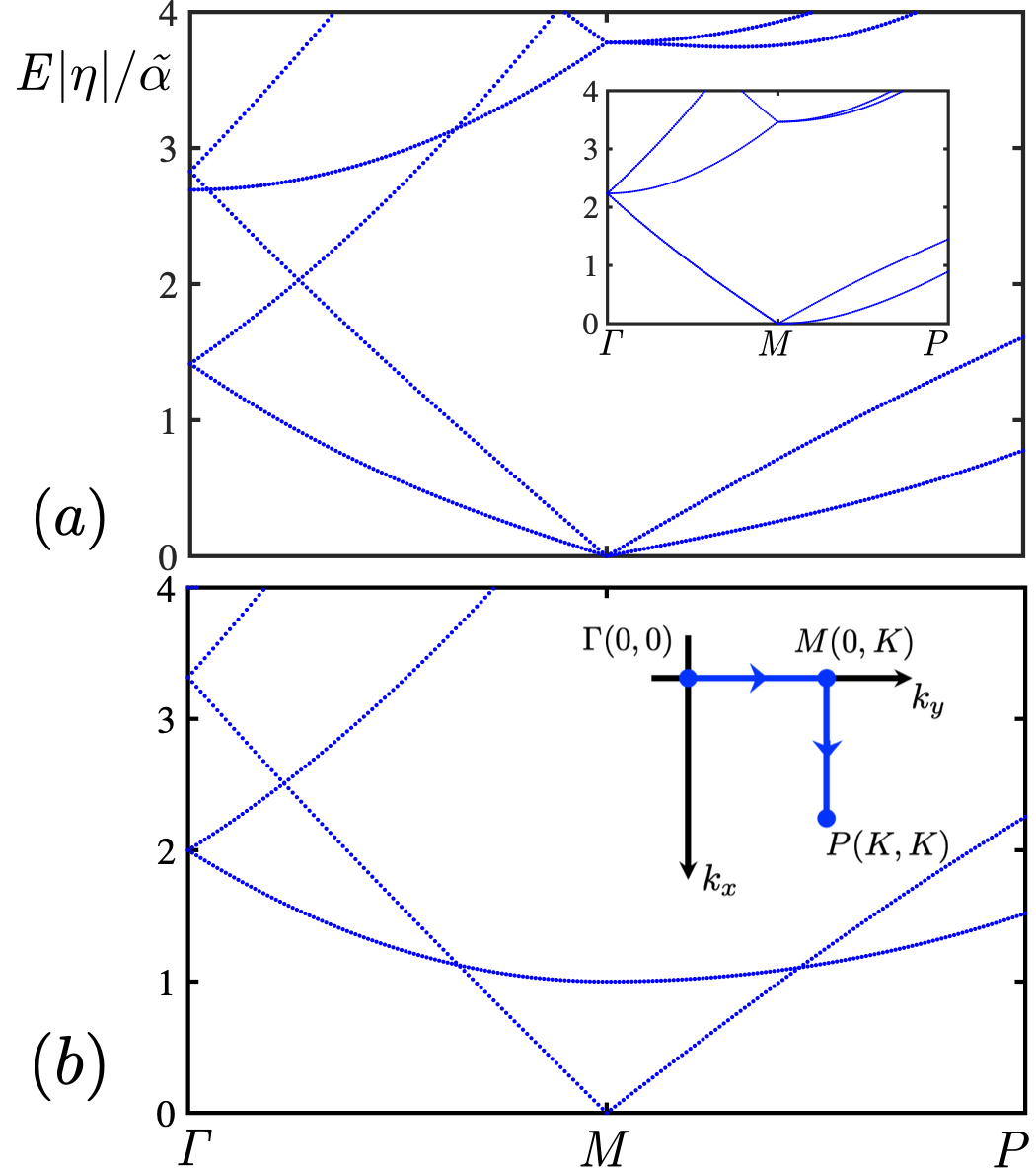}
	\caption{(color online) Energy-momentum dispersions of the 
low-energy collective modes in the helicoidal excitonic condensate phases. 
The dispersions are plotted along the high symmetric line (shown in the inset of (b)). 
The unit of the energy axis is $\tilde{\alpha}/|\eta|$ where 
$\tilde{\alpha} \equiv \alpha-\frac{2}{g}$. 
(a) The dispersion for $h<h_c$ where $\tilde{\alpha}=1,|\eta|=1,\lambda=1,D=2,h=1.5$ 
(inset is for $h=0$). (b) The dispersion for 
$h>h_c$, where $\tilde{\alpha}=1,|\eta|=1,\lambda=1,D=2,h=3$.} 
	\label{fig:2}
\end{figure}

As suggested by the Berry phase term in the effective action, ${\bm \phi}^{\dagger}\partial_{\tau} {\bm \phi} = 
i{\bm \phi}^{\dagger}\partial_{t} {\bm \phi}$,  $\delta {\bm \phi}({\bm x})$ and $\delta {\bm \phi}^{\dagger}({\bm x})$ 
are nothing but (Holstein-Primakov) boson annihilation and creation operator respectively. Accordingly, the linearized EOMs 
thus obtained must reduce to a generalized eigenvalue problem with bosonic 
Bogoliubov de-Gennes (BdG) Hamiltonian~\cite{colpa78};
\begin{align}
|\eta| i \frac{\partial}{\partial t} 
\left(\begin{array}{c}
\delta {\bm \phi} ({\bm x}) \\
\delta {\bm \phi}^{\dagger} ({\bm x}) \\
\end{array}\right) = 
{\bm \tau}_3 \hat{\bm H}_{\rm BdG}({\bm \nabla},{\bm x})  
\left(\begin{array}{c}
\delta {\bm \phi}({\bm x}) \\
\delta {\bm \phi}^{\dagger}({\bm x}) \\
\end{array} \right). \label{bdg0}
\end{align}
Here ${\bm \tau}_3$ is the 2 by 2 diagonal Pauli matrix in the particle-hole space, that  
takes $+1$ for the annihilation and $-1$ for the creation operator. 
$\hat{\bm H}_{\rm BdG}({\bm \nabla},{\bm x})$ is a 6 by 6 matrix-formed differential operators, that are hermitian,
$\hat{\bm H}^{\dagger}_{\rm BdG}({\bm \nabla},{\bm x})=\hat{\bm H}_{\rm BdG}(-{\bm \nabla},{\bm x})$. 
For $h\le h_c$, the BdG Hamiltonian thus obtained takes the following explicit form, 
\begin{widetext}
\begin{align}
{\bm H}_{\rm BdG} = \left(\begin{array}{cccccc}
-\alpha_{\nabla} & 0 & D \partial_x & - 2|\gamma| \rho^2 F_{0} & 0 & 0 \\
0 & -\alpha_{\nabla} & D \partial_y - ih & 0 & C_y &  S_y \\
- D\partial_x & - D \partial_y + ih & -\alpha_{\nabla}  & 0 &  S_y & -C_y \\
- 2|\gamma| \rho^2 F^{*}_{0} & 0 & 0 & -\alpha_{\nabla} & 0 & D\partial_x \\
0 & C^*_y & S^{*}_y & 0 & -\alpha_{\nabla}   & D \partial_y + ih \\
0 & S^{*}_y & - C^{*}_y & - D\partial_x & - D\partial_y -ih & -\alpha_{\nabla}  \\
\end{array} \right).  \label{bdg1}
\end{align}
For $h\ge h_c$, the BdG Hamiltonian is given by 
\begin{align}
{\bm H}_{\rm BdG} = \left(\begin{array}{cccccc}
-\alpha^{\prime}_{\nabla} & 0 & D \partial_x & 0 & 0 & 0 \\
0 & -\alpha^{\prime}_{\nabla} & D \partial_y + ih - i2\zeta & 0 & \zeta e^{2iKy} 
 & i \zeta e^{2iKy}  \\
- D\partial_x & - D \partial_y - ih + i 2\zeta & -\alpha^{\prime}_{\nabla}  & 0 & i \zeta e^{2iKy}  &  - \zeta e^{2iKy}  \\
0 & 0 & 0 & -\alpha^{\prime}_{\nabla}  & 0 & D\partial_x \\
0 & \zeta e^{-2iKy}  & - i\zeta e^{-2iKy}  & 0 & -\alpha^{\prime}_{\nabla}  & D \partial_y - ih + i2 \zeta  \\
0 & - i \zeta e^{-2iKy} & - \zeta e^{-2iKy} & - D\partial_x 
& - D\partial_y + ih -i 2\zeta & -\alpha^{\prime}_{\nabla} \\
\end{array} \right) , \label{bdg2}
\end{align}
\end{widetext}
with 
\begin{align}
\zeta & \equiv 2|\gamma| {\rho^{\prime}}^2, \label{zeta} \\
\alpha_{\nabla} & \equiv \alpha -\frac{2}{g} - 4|\gamma| \rho^2  + \lambda {\nabla}^2, \label{a1} \\
\alpha^{\prime}_{\nabla} &\equiv 
\alpha -\frac{2}{g} - 4|\gamma| {\rho^{\prime}}^2  + \lambda {\nabla}^2, \label{a2} \\ 
F_{0} &= \cos2\theta + i \sin 2\theta \cos\nu, \label{f0} \\ 
C_{y} &\equiv \frac{h_c}{2}  
\big( e^{i2Ky} F_{+} + e^{-i2Ky} F_{-}\big),  \label{cy} \\ 
S_{y} & \equiv i \frac{h_c}{2}  
\big( e^{i2Ky} F_{+}  - e^{-i2Ky} F_{-}\big),  \label{sy} \\
F_{\pm} &\equiv \cos^2\theta - \sin^2 \theta e^{\mp i2\nu} 
+i \sin 2\theta e^{\mp i \nu}. \label{fpm} 
\end{align}
$K$, $\rho$, $\rho^{\prime}$ and 
$h_c$ are already defined by Eqs.~(\ref{hs-0-c},\ref{hs-2-b},\ref{satu}) respectively. 
$\theta$ and $\nu$ in Eqs.~(\ref{f0},\ref{fpm}) must satisfy Eq.~(\ref{hs-1-c}). 
Note that the two BdG Hamiltonians beome identical to each other at 
$h=h_c$, where $F_{0}=0$, $F_{+}=2$, $F_{-}=0$. Using bosonic Bogoliubov 
transformations~\cite{colpa78}, we diagonalize these Hamiltonians in the momentum space, 
to obtain energy-momentum dispersions of the low-energy 
collective excitations in the helicoidal excitonic phases (Appendix~C). Fig.~\ref{fig:2} 
shows the dispersions along the high symmetric line in the momentum space 
for $h=0$, $h<h_c$ and $h>h_c$ respectively.  

\begin{figure}[t]
	\centering
	\includegraphics[width=0.8\linewidth]{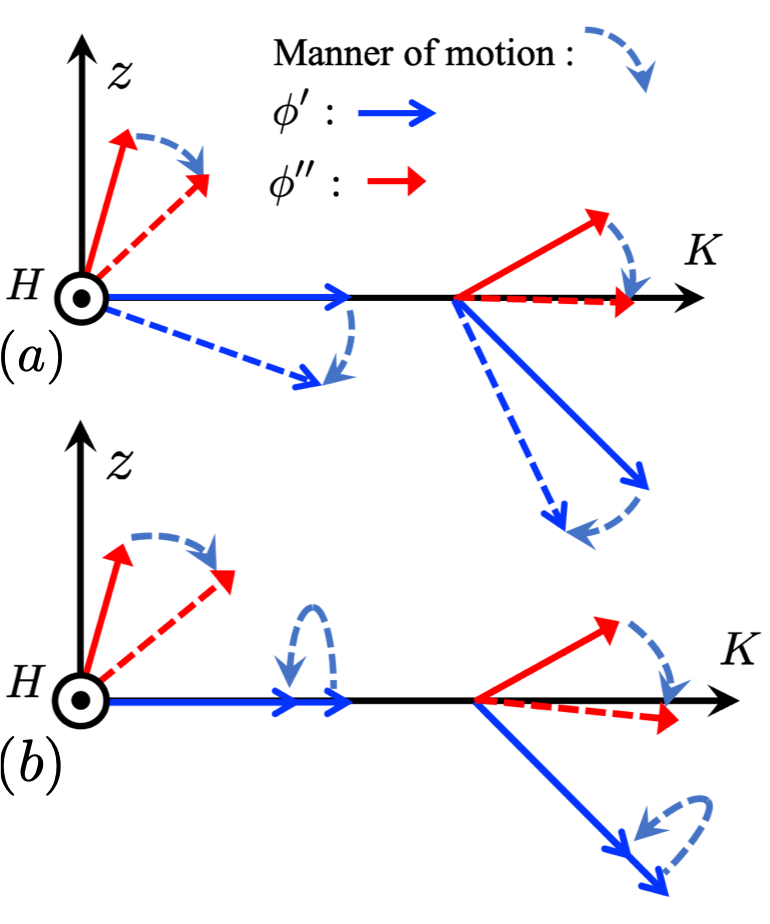}
	\caption{(color online) Schematic pictures of (a) translational mode and (b) spin rotational mode} 
	\label{fig:3}
\end{figure}

The helicoidal condensate phase at $h\le h_c$ has two gapless Goldstone modes around 
${\bm k}=(0,K)$; translational mode and spin rotational mode (Fig.~\ref{fig:2}(a)). These 
two result from the spontaneous symmetry breakings of the continuous symmetries;   
translational symmetry and a combined symmetry of the relative gauge symmetry and the spin 
rotation symmetry respectively. The translational mode at the gapless point induces a 
simultaneous rotation of ${\bm \phi}^{\prime}$ and ${\bm \phi}^{\prime\prime}$ 
by the same angle around the $yz$ plane (Fig.~\ref{fig:3}(a)). The spin rotational 
mode induces a change of the amplitudes of  ${\bm \phi}^{\prime}$ and ${\bm \phi}^{\prime\prime}$ 
as well as the relative angle between these two vectors (Fig.~\ref{fig:3}(b)), but it does not change 
the total amplitude of the spin-1 exciton field, 
$|{\bm \phi}|^2 \equiv |{\bm \phi}^{\prime}|^2 + |{\bm \phi}^{\prime\prime}|^2$. When the 
field is above the critical field ($h\ge h_c$), the relative angle between the real and imaginary part is locked 
to $\pi/2$ by the large in-plane field; the angle between these two are fully saturated (`fully saturated phase'). 
Accordingly, the spin rotational mode acquires a finite mass and only the translational mode forms a gapless 
dispersion at ${\bm k}=(0,K)$ (See Fig.~\ref{fig:2}(b)). 

Finally, let us give several remarks on light scatterings for probing these low-energy collective modes  
in the EHDL system. In the layered heterostructure, the electron and hole layers are physically well 
separated by an intermediate separation layer with its thickness being typically 10 nm~\cite{wu19}. 
Meanwhile, single exciton is a pair of the electron creation and the hole creation. Thus, generally, 
the external electromagnetic field can couple with the exciton in the EHDL system 
only through 2-excitons process, 4-excitons process, 
and so on; one photon causes at least a pair of exciton creation and exciton annihilation in the EHDL system. 
A calculation in appendix D shows that the (static) helicoidal excitonic order parameter quadratically induces 
a uniform change of the electron density in the electron layer, $\delta \rho^{e}(\bm x) \equiv 
\delta \langle a^{\dagger}_{\alpha}({\bm x})a_{\alpha}({\bm x})\rangle \sim \rho^2$. 
This suggests that photon can couple {\it electrically} with the low-energy collective modes in the helicoidal 
excitonic condensate in EHDL system.

\section{experimental probes}  
In the presence of a small Dirac term in the hole layer $(\Delta_h \ne 0)$, the helicoidal texture of the spin-1 
exciton condensate induces a helicoidal texture of local {\it magnetic} moment in the electron layer, whose 
spatial pitch is $1/2K$ instead of $1/K$. At the zero magnetic field, however, the helicoid structure given by 
Eq.~(\ref{hs-0-a}) is symmetric under the time-reversal symmetry combined with a gauge transformation 
for the electron or hole band. Thus, the local magnetic moment in each layer is quenched at $H=0$.

The helicoidal {\it magnetic} texture appears when the in-plane Zeeman field $H$ is applied to the helicoidal excitonic 
phase. Thereby, the local magnetic moment as well as the spin-1 exciton field   
rotate around the in-plane Zeeman field; 
\begin{align}
&{\bm m}^e({\bm x}) \equiv \langle a^{\dagger}_{\alpha}({\bm x})[{\bm \sigma}]_{\alpha\beta} a_{\beta}({\bm x}) \rangle \nonumber \\ 
&= A(H) \big(\cos^2\theta \cos (2Ky) + \sin^2 \theta \cos(2Ky-2\nu) \big) \hat{\bm e}_y -  \nonumber \\
& \hspace{-0.1cm} B(H) \big(\cos^2\theta \sin (2Ky) + \sin^2 \theta \sin(2Ky-2\nu) \big) \hat{\bm e}_z + \cdots. \label{mag}
\end{align} 
Here $A(H)$ and $B(H)$ are real-valued and odd functions in the magnetic field $H$; 
$A(-H)=-A(H)$, $B(-H)=-B(H)$. `$\cdots$' in the right-hand side 
denotes the higher order harmonic contributions ($4Ky$, $6Ky$, $\cdots$ components). 
In the leading order in small $\Delta_{h}$, $A$ and $B$ are proportional to 
$\rho^2 \Delta_h$ (Appendix D).  

Having a finite out-of-plane (${\bm e}_z$-component) magnetization, the spatial magnetic texture 
given in Eq.~(\ref{mag}) could be experimentally seen by magnetic optical measurements~\cite{hubert98}.  
For example, when the electron layer is sufficiently reflective, a spatial map of the magnetic Kerr rotational 
angle in the two-dimensional layer must show a stripe structure. According to our theory prediction, 
the magnetic stripe appears in parallel to the in-plane Zeeman field, and it disappears when the field 
is set zero. The Kerr rotation angle changes its sign when the in-plane field is reversed.   

\section{conclusion and discussion} 
A recent transport experiment on a strained layered InAs/AlSb/GaInSb heterostructure reports resistive 
signatures of the excitonic coupling at low temperature around the charge neutrality line of the 
BEC regime~\cite{wu19}.  
In this paper, we show that due to the large Rashba interaction in the electron layer, energy 
degeneracy among three spin-1 (spin-triplet) 
exciton bands are lifted at finite momentum. On lowering the temperature or on changing 
the charge state energy, the lowest spin-1 exciton band can undergo the BEC at finite momentum, 
resulting in a helicoidal structure of the spin-1 exciton field (helicoidal excitonic condensate). The 
helicoidal plane of the spin-1 exciton can be controlled by the in-plane Zeeman field. 

Based on the linearized coupled EOMs of the spin-1 exciton, we calculate 
momentum-energy dispersions of the low-energy collective modes in the helicoidal excitonic phase. 
For future possible light scattering experiments, we show that these low-energy modes can couple electrically 
with one photon through the two-excitons processes. We also demonstrate that due to small Dirac 
term in the heavy hole layer, the helicoidal structure of the spin-1 exciton condensate 
results in a helicoidal {\it magnetic} structure in the electron layer. 
Having a finite out-of-plane magnetization, the helicoidal magnetic structure could 
be visualized by a spatial map of the magnetic Kerr rotation angle in the two-dimensional layer. Our 
theory predicts that the magnetic stripes appear in parallel to the in-plane Zeeman field and it 
disappears at the zero Zeeman field.    

In a Coulomb coupled EHDL system without the SOI, the spin triplet (spin-1) excitons and 
spin singlet (spin-0) exciton are energetically degenerate due to the independent spin rotations 
of electron spin and hole spin~\cite{zhu96}. Meanwhile, the analyses in this paper has ignored a 
coupling between the spin-0 exciton and spin-1 excitons. The coupling {\it does} exist 
at the first order in the Rashba term in the electron layer $\xi_e$ as well as at the 
first order in the in-plane Zeeman field $H$;
\begin{align}
S &=  \cdots - D \int d{\bm x} \!\ \Big[\hat{\bm e}_y \cdot
\big({\bm \phi}^{\dagger} \times \partial_x {\bm \phi}\big)  
- \hat{\bm e}_x \cdot \big({\bm \phi}^{\dagger} \times \partial_y {\bm \phi}\big) \Big] \nonumber \\
& + D \int d{\bm x} \!\ \Big[ 
\phi^{\dagger}_x i\partial_y \phi_0 + \phi^{\dagger}_0 i\partial_y \phi_x 
- \phi^{\dagger}_y i\partial_x \phi_0 - \phi^{\dagger}_0 i\partial_x \phi_y \Big] 
\nonumber \\
 & \hspace{-0.7cm} 
+ ih \int d{\bm x} \!\ \hat{\bm e}_H \cdot \big({\bm \phi}^{\dagger}\times {\bm \phi}\big) 
+ h^{\prime} \int d{\bm x} \!\ \hat{\bm e}_H \cdot 
\big({\bm \phi}^{\dagger} \phi_0 + \phi^{\dagger}_0 {\bm \phi} \big),  
\label{additional}
\end{align}
Here ${\bm \phi}$ and $\phi_{0}$ denote the spin-1 and spin-0 exciton fields 
respectively. The first and third terms are nothing but the last three terms in Eq.~(\ref{action}). 
$h$ and $h^{\prime}$ are proportional to the in-plane field $H$, $h \ne h^{\prime}$. 
At the quadratic level of the effective action at the zero field 
($h=h^{\prime}=0$), the four-fold degenerate exciton bands 
at the zero momentum (one spin-0 and three spin-1) are split into two doubly 
degenerate exciton bands at finite momentum ${\bm k}$;
\begin{widetext}
\begin{align}
H_{{\rm EX}} 
= \sum_{\bm k}
\left(\begin{array}{cccc}
\phi^{\dagger}_0({\bm k}) & \phi^{\dagger}_{x}({\bm k}) & \phi^{\dagger}_y({\bm k})
 & \phi^{\dagger}_z({\bm k}) \\
\end{array}\right) \left(\begin{array}{cccc} 
\beta_{\bm k} & - k_y D & k_x D & 0 \\
- k_y D & \beta_{\bm k} & 0 & ik_x D \\
k_x D & 0 & \beta_{\bm k} & ik_y D \\
0 & -ik_x D & -ik_y D  & \beta_{\bm k} \\
\end{array}\right) \left(\begin{array}{c}
\phi_0({\bm k}) \\
\phi_{x}({\bm k}) \\
\phi_{y}({\bm k}) \\
\phi_{z}({\bm k}) \\
\end{array}\right), \label{EX} 
\end{align}
\end{widetext} 
with $\beta_{\bm k} \equiv -\alpha+2/g+\lambda k^2>0$. The upper two 
exciton bands have an energy of $\beta_{\bm k} + D|{\bm k}|$, and the lower two  
exciton bands have an energy of $\beta_{\bm k} - D|{\bm k}|$. One of the lower two bands is 
a mixture of purely three spin-1 excitons, $\phi_{z} + i (\hat{k}_x \phi_x + \hat{k}_y \phi_y)$, 
whose BEC induces the helicoidal excitonic phase [discussed in 
this paper]. On the one hand, the other of the lower two bands is a mixture of the spin-0 
and spin-1 exciton bands, $\phi_{0} -  (-\hat{k}_y \phi_x + \hat{k}_x \phi_y)$,  
whose BEC induces an in-plane collinear texture of the spin-1 exciton field. The collinear texture 
and the helicoidal texture are energetically degenerate at the zero Zeeman field in the absence of 
the Dirac term in hole layer ($\Delta_h=H=0$). The energy degeneracy is due to the $\pi$ spin-rotation 
around the $z$-axis {\it only} in the hole band; 
${\bm b}^{\dagger}_{\bm k} \rightarrow {\bm b}^{\dagger}_{\bm k}{\bm \sigma}_z$ 
in Eqs.~(\ref{hami1},\ref{hami2}). The degeneracy is lifted by the Dirac term in the hole band 
$\Delta_h$ as well as the in-plane Zeeman field. In the presence of these perturbations, a mixture 
of these two textures will be selected as a true classical ground state. By construction, the mixture 
has lower symmetries than the helicoidal excitonic phase discussed in this paper 
and thereby it has essentially the same physical response as the helicoidal phase [Sec. VI and Sec. V]. 
Nonetheless, detailed physical properties of the mixed phase need more theoretical studied and will be 
discussed elsewhere.  

  
\section*{ACKNOWLEDGMENTS}

RS thank Rui-Rui Du for helpful information and discussion. This work was 
supported by NBRP of China (Grant No. 2014CB920901, Grant No. 2015CB921104 and 
Grant No. 2017A040215). 

\appendix 
\section{derivation of the effective action}
In this appendix, we derive an effective $\phi^4$ action for the spin-triplet (spin-1) exciton field  
in the presence of the SOI and the Zeeman field. We begin with the 
partition function $Z$ for Eq.~(\ref{hami1},\ref{hami2},\ref{hami3}) 
with $\Delta_{h}=0$, 
\begin{align}
Z = \int \prod {\cal D}{\bm a}^{\dagger}_{k} {\cal D}{\bm a}_{k}
{\cal D}{\bm b}^{\dagger}_{k} {\cal D}{\bm b}_{k} 
\exp \big[- S[{\bm a},{\bm b}] \big] \nonumber 
\end{align}
where the effective action $S$ is given by
\begin{widetext}
\begin{align}
S[{\bm a},{\bm b}] = \sum_{k} \left(\begin{array}{cc} 
{\bm a}^{\dagger}_{k} & {\bm b}^{\dagger}_{k} \\
\end{array}\right) \!\  
\big( G^{-1}_{0}(k) + G^{-1}_{\rm R}(k) + G^{-1}_{H}(k)\big) \!\ \left(\begin{array}{c}
{\bm a}_{k} \\
{\bm b}_{k} \\
\end{array}\right) - \frac{g}{2} \sum_{k} \sum_{m=0,x,y,z} {\cal O}^{\dagger}_m(k) {\cal O}_m(k) \Big]. 
\end{align}
Here non-interacting temperature Green function, Rashba and Zeeman field parts take forms of,
\begin{align}
G^{-1}_0(k) & \equiv \left(\begin{array}{cc} 
(-i\omega_n+{\cal E}_{a}({\bm k}) -\mu) {\bm \sigma}_0 & \\
& (-i\omega_n+{\cal E}_{b}({\bm k}) -\mu) {\bm \sigma}_0 \\
\end{array}\right), \label{g0} \\
G^{-1}_{\rm R}(k) &\equiv \left(\begin{array}{cc} 
\xi_e( k_y {\bm \sigma}_x  - k_x {\bm \sigma}_y ) & \\
& 0 \\
\end{array}\right),  \ \  G^{-1}_{H}(k)  \equiv \left(\begin{array}{cc} 
H {\bm \sigma}_{H} & \\
& H {\bm \sigma}_{H} \\
\end{array}\right), \label{grh} 
\end{align} 
\end{widetext}
with ${\cal E}_{a}({\bm k}) \equiv \hbar^2 {\bm k}^2/2m_e - E_g$, 
${\cal E}_{b}({\bm k})\equiv - \hbar^2 {\bm k}^2/2m_h + E_g$, and 
$k\equiv (i\omega_n,{\bm k})$. Note that we used the following Fourier transformation 
for ${\bm a}({\bm x},\tau)$, ${\bm b}({\bm x},\tau)$ and ${\cal O}({\bm x},\tau)$, 
\begin{align}
f({\bm x},\tau) = \frac{1}{\sqrt{\beta V}} \sum_k e^{i{\bm k}{\bm x}-i\omega_n \tau} 
f(k).  
\end{align} 
$\beta$ is an inverse temperature, $\beta\equiv 1/(k_{\rm B}T)$, $i\omega_n=(2n+1)\pi/\beta$ for 
${\bm a}$ and ${\bm b}$, $i\omega_n=2n\pi/\beta$ for ${\cal O}$, and $\sum_k\equiv \sum_{\bm k}
\sum_{\omega_n}$. 

A decomposition of the interaction by the Stratonovich Hubbard (SH) variables ${\bm \phi}(k)$~\cite{fradkin91} 
gives out a quadratic form of the ${\bm a}$ and ${\bm b}$ fields,
\begin{align}
&\exp \Big[\frac{g}{2} \sum_k \sum^z_{m=0} {\cal O}^{\dagger}_m(k) {\cal O}_m(k)\Big]  \nonumber \\
& \hspace{0.5cm} 
= \int {\cal D}{\bm \phi}^{\dagger} {\cal D}{\bm \phi} \!\ \exp \Big[-\sum_{k} \frac{2}{g} |{\bm \phi}_{k}|^2 \nonumber \\
& \hspace{1cm} 
+ \sum_k \big({\bm \phi}^{\dagger}(k) \cdot {\cal O}(k) + {\cal O}^{\dagger}(k) \cdot {\bm \phi}(k)\big) \Big]. 
\end{align}
A gaussian integration over the ${\bm a}$ and ${\bm b}$ fields leads to a functional of the SH variables. A Taylor 
expansion of the functional with respect to the SH variables gives,   
\begin{align}
& Z = \int {\cal D}{\bm \phi}^{\dagger} {\cal D}{\bm \phi} \!\ \exp \Big[-\sum_{k} \frac{2}{g} |{\bm \phi}_{k}|^2 \nonumber \\
& \hspace{0.8cm} + {\rm Tr} \big[{\bm 1} + G_0(k) \!\ G^{-1}_{\rm R}(k) \!\ \delta_{k,k'} \nonumber \\
&\hspace{1.2cm} + G_0(k) \!\ G^{-1}_{H}(k) \!\ \delta_{k,k'}  - G_0(k) \!\ {\Phi}_{q} \!\ \delta_{k,k+q} \big] \Big] \nonumber \\
&  = \int {\cal D}{\bm \phi}^{\dagger} {\cal D}{\bm \phi} \!\ \exp \Big[-\sum_{k} \frac{2}{g} |{\bm \phi}_{k}|^2 \nonumber \\
& \hspace{0.8cm} + {\rm Tr} \big[G_0 G^{-1}_{\rm R} G_0 {\Phi} G_0 {\Phi} + 
G_0 G^{-1}_{H} G_0 {\Phi} G_0 {\Phi}  \nonumber \\
& \hspace{1.2cm}  - \frac{1}{2} G_0 {\Phi} G_0 {\Phi} - \frac{1}{4} 
 G_0 {\Phi} G_0 {\Phi} G_0 {\Phi} G_0 {\Phi}\big] \Big], \label{exp}
\end{align}
where 
\begin{align}
{\Phi}_{q} = \frac{1}{\sqrt{\beta V}} \left(\begin{array}{cc} 
0  &\sum_{m} \phi_{m}(-q) {\bm \sigma}_{m} \\ 
\sum_{m} \phi^{\dagger}_{m}(q) {\bm \sigma}_{m} & 0 \\
\end{array}\right).  
\end{align} 
In the expansion, we took into account only the first order in the small $\xi_e$ and $H$. We also 
expand Eq.~(\ref{exp}) in small $q\equiv (i\epsilon_n,{\bm q})$, to keep up to 
${\cal O}({\bm q}^2,i\epsilon_n)$ in ${\rm Tr} [G_0 {\Phi} G_0 {\Phi}]$, 
up to ${\cal O}({\bm q}^1,i\epsilon_n^0)$ in ${\rm Tr} [G_0 G^{-1}_{\rm R} G_0 {\Phi} G_0 {\Phi}]$, 
and up to ${\cal O}({\bm q}^0,i\epsilon_n^0)$ in the other terms. This gives Eq.~(\ref{action}) as the effective 
action for the spin-1 exciton field. The coefficients in Eq.~(\ref{action}) are calculated in the followings,
\begin{align}
-\alpha &+ \lambda {\bm q}^2 + \cdots = \frac{2}{V} \sum_{\bm k} \frac{n_{\rm F}({\cal E}_{a}({\bm k}))
-n_{\rm F}({\cal E}_{b}({\bm k}+{\bm q}))}{{\cal E}_{a}({\bm k})-{\cal E}_{b}({\bm k}+{\bm q})}, \nonumber \\
\eta &=  -\frac{2}{V} \sum_{\bm k} \bigg\{ \frac{n_{\rm F}({\cal E}_{b}({\bm k}))
-n_{\rm F}({\cal E}_{a}({\bm k}))}{({\cal E}_{a}({\bm k})-{\cal E}_{b}({\bm k}))^2} \nonumber \\
& \hspace{-0.5cm} 
- \frac{\beta}{2+2\cosh (\beta({\cal E}_{a}({\bm k})-\mu))} \frac{1}{{\cal E}_{a}({\bm k})-{\cal E}_{b}({\bm k})} \bigg\} <0,  
\nonumber  \\ 
\gamma &= - \frac{1}{V} \sum_{\bm k} \bigg(\frac{1}{{\cal E}_{a}-{\cal E}_{b}}\bigg)^2 \bigg\{ 2 
\frac{n_{\rm F}({\cal E}_b) - n_{\rm F}({\cal E}_a)}{{\cal E}_{a}-{\cal E}_{b}}  \nonumber \\
& \hspace{-0.8cm} - \frac{\beta}{2+2\cosh (\beta({\cal E}_{b}-\mu))} - \frac{\beta}{2+2\cosh (\beta({\cal E}_{a}-\mu))} \bigg\}  
<0,  
\nonumber 
\end{align}
\begin{align}
D &= -\frac{\xi_e}{V} \sum_{\bm k} \frac{\hbar^2 k^2_x}{m_h} 
\bigg(\frac{1}{{\cal E}_{a}-{\cal E}_{b}}\bigg)^2 \bigg\{ 4 
\frac{n_{\rm F}({\cal E}_b) - n_{\rm F}({\cal E}_a)}{{\cal E}_{a}-{\cal E}_{b}}  \nonumber \\
& \hspace{-0.5cm} - \frac{\beta}{1+\cosh (\beta({\cal E}_{b}-\mu))} - \frac{\beta}{1+\cosh (\beta({\cal E}_{a}-\mu))} \bigg\},  
\nonumber \\
h &= \frac{2H}{V} \sum_{\bm k} \frac{1}{{\cal E}_{a}-{\cal E}_{b}} \bigg\{ 2 
\frac{n_{\rm F}({\cal E}_a) - n_{\rm F}({\cal E}_b)}{{\cal E}_{a}-{\cal E}_{b}}  \nonumber \\
& \hspace{-0.5cm} + \frac{\beta}{2+2\cosh (\beta({\cal E}_{b}-\mu))} + \frac{\beta}{2+2\cosh (\beta({\cal E}_{a}-\mu))} \bigg\}.  
\nonumber 
\end{align}
From these expressions, we can see that both $\alpha$ and $\lambda$ are positive. Since $D$ and $h$ are proportional to 
$\xi_e$ and $H$ respectively, we can assume that $D$ and $h$ are positive without loss of generality.  

\section{minimization of the classical action}
In this appendix, we minimize the classical energy in Eq.~(\ref{action}) with respect to the real and imaginary parts of the spin-1 
exciton field, ${\bm \phi}={\bm \phi}^{\prime}+i{\bm \phi}^{\prime\prime}$. Take ${\bm \phi}^{\prime}=A{\bm n}^{\prime}$ 
and ${\bm \phi}^{\prime\prime}=B{\bm n}^{\prime\prime}$ with unit vectors ${\bm n}^{\prime}$ and  ${\bm n}^{\prime\prime}$
; $|{\bm n}^{\prime}|=|{\bm n}^{\prime\prime}|=1$. In the absence of the SOI, the exciton field 
takes a spatially uniform solution, because $\lambda>0$. Thereby, the real and imaginary parts are parallel to each other in the spin space;
\begin{align}
{\bm \phi}^{\prime}_{c}+i{\bm \phi}^{\prime\prime}_c = \frac{1}{2|\gamma|} \Big(\alpha-\frac{2}{g}\Big) e^{i\theta} {\bm n}, 
\end{align} 
The solution breaks the two global symmetries. The U(1) symmetry associated with $\theta$ is nothing but 
a difference between the U(1) gauge degree of freedom of the electron and that of the hole. 
The SO(3) symmetry associate with ${\bm n}$ represents the global spin rotational symmetry. 
In the following, we study how the uniform solution would be 
deformed in the presence of finite SOI ($D\ne 0$). 
\subsection{$H=0$ and $D \ne 0$} 
In the presence of the SOI, the spin-1 exciton field takes a spatially dependent solution. Spatial gradients 
of the amplitudes, $A$ and $B$, do not lower the SOI energy. Accordingly, without loss of generality, we can 
assume that the amplitudes are spatially uniform and the unit vectors depend on the space coordinate ${\bm x}$;
\begin{align}
{\bm \phi}^{\prime} &= A {\bm n}^{\prime}({\bm x}), \label{asp1} \\
{\bm \phi}^{\prime\prime} &= B {\bm n}^{\prime\prime}({\bm x}). \label{asp2} 
\end{align}
This gives the following functional for the classical action, where $V$ and $\beta$ are the volume of the system and 
the inverse temperature respectively;
\begin{align}
&\frac{1}{\beta V} S[A,B,{\bm n}^{\prime}({\bm x}),{\bm n}^{\prime\prime}({\bm x})]  \nonumber \\
& \ \ \ = - \big(\alpha -\frac{2}{g} \big) \big(A^2+B^2 \big) - \gamma \big( A^4 + B^4 + 6 A^2 B^2 \big)  \nonumber \\
& \ \ -\frac{A^2}{V} C_{1}[{\bm n}^{\prime}] - \frac{B^2}{V} C_{1}[{\bm n}^{\prime\prime}] 
+ 4\gamma \frac{A^2 B^2}{V} C_{2} [{\bm n}^{\prime},{\bm n}^{\prime\prime}]. \label{action01}  
\end{align} 
Here $C_1$ are $C_2$ are functionals of the unit vectors,  
\begin{align}
C_{1}[{\bm n}^{\prime}] \equiv & \int d{\bm x} \Big[ - \lambda |{\nabla} {\bm n}^{\prime}|^2 \nonumber \\
& \hspace{-1.2cm} + D\big({\bm e}_y \cdot ({\bm n}^{\prime}\times \partial_x {\bm n}^{\prime}) 
-  {\bm e}_x \cdot ({\bm n}^{\prime}\times \partial_y {\bm n}^{\prime}) \big) \Big], \label{c1} \\
C_{2}[{\bm n}^{\prime},{\bm n}^{\prime\prime}] & \equiv 
\int d{\bm x} \!\ \big({\bm n}^{\prime} \cdot {\bm n}^{\prime\prime}\big)^2. \label{c2} 
\end{align}
For the later convenience, let us rotate ${\bm n}^{\prime}$ and ${\bm n}^{\prime\prime}$ by $\pi/2$ around 
the $z$-axis;
\begin{align}
{\bm n}^{\prime(\prime\prime)}_{\rm new}({\bm x}) = \left(\begin{array}{ccc}
0 & -1 & 0 \\
1 & 0 & 0 \\
0 & 0 & 1 \\
\end{array}\right) {\bm n}^{\prime(\prime\prime)}_{\rm old}({\bm x}). \label{rotate} 
\end{align}
In the rotated frame, $C_{1}$ and $C_2$ are given by 
\begin{align}
C_{1}[{\bm n}^{\prime}] = & \int d{\bm x} \Big[ - \lambda |{\bm \nabla} {\bm n}^{\prime}|^2 
+ D {\bm n}^{\prime}\cdot ({\bm \nabla} \times {\bm n}^{\prime}) \Big] \label{c1r} \\
C_{2}[{\bm n}^{\prime},{\bm n}^{\prime\prime}] &= 
\int d{\bm x}  \!\ \big({\bm n}^{\prime}\cdot {\bm n}^{\prime\prime}\big)^2. \label{c2r} 
\end{align}
with ${\bm \nabla} \equiv (\partial_x,\partial_y,0)$. 

In the following, we first minimize the classical action for fixed $A$ and $B$ (Eq.~(\ref{action01})). 
To this end, we have only to maximize $C_1[{\bm n}^{\prime}]$, 
$C_{1}[{\bm n}^{\prime\prime}]$ and $C_{2}[{\bm n}^{\prime},{\bm n}^{\prime\prime}]$ 
with respect to ${\bm n}^{\prime}$ and ${\bm n}^{\prime\prime}$, because $\gamma<0$. 
These three functionals can be simultaneously maximized. 

To see this, let us first maximize $C_{1}[{\bm n}]$ under the normalization condition 
of $|{\bm n}({\bm x})|=1$ for any ${\bm x}$. In the momentum-space representation, the Fourier series 
of ${\bm n}({\bm x})$ comprises of two real-valued vectors, ${\bm \alpha}_{\bm k}$ and ${\bm \beta}_{\bm k}$, 
\begin{align}
{\bm n}({\bm x}) &= \sum_{\bm k} e^{i{\bm k}{\bm x}} \frac{1}{2} \big({\bm \alpha}_{\bm k}+ i {\bm \beta}_{\bm k}\big) 
\nonumber \\
&= \sum_{k_x>0} \big(\cos({\bm k}{\bm x}) {\bm \alpha}_{\bm k} - \sin({\bm k}{\bm x}) {\bm \beta}_{\bm k} \big),   
\label{fourier}
\end{align}
with ${\bm \alpha}_{\bm k}= {\bm \alpha}_{-{\bm k}}$ and ${\bm \beta}_{\bm k}=-{\bm \beta}_{-{\bm k}}$. 
In terms of these vectors, $C_1[{\bm n}]$ takes a form of 
\begin{align}
C_{1}[{\bm n}] = V \sum_{k_x>0} \Big[-\frac{\lambda}{2} \big(|{\bm \alpha}_{\bm k}|^2 + |{\bm \beta}_{\bm k}|^2
\big) k^2 + D{\bm k}\cdot \big({\bm \alpha}_{\bm k} \times {\bm \beta}_{\bm k}\big) \Big]. \label{c1f}
\end{align}
The normalization condition imposes a global constraint onto the Fourier series;
\begin{align}
\frac{1}{V} \int d{\bm x} \!\ |{\bm n}({\bm x})|^2 &= \frac{1}{2} \sum_{k_x>0} 
\big(|{\bm \alpha}_{\bm k}|^2 + |{\bm \beta}_{\bm k}|^2\big) \nonumber \\ 
&\equiv \frac{1}{2} \sum_{k_x>0} w^2_{\bm k} = 1. \label{global} 
\end{align}
Under Eq.~(\ref{global}), Eq.~(\ref{c1f}) is maximized by 
\begin{align}
{\bm n}({\bm x}) = \sum^{k_x>0}_{|{\bm k}|=D/2\lambda} \frac{w_{\bm k}}{\sqrt{2}} 
\big(\cos({\bm k}{\bm x}) \hat{\bm k}_{\perp,1} - \sin({\bm k}{\bm x}) \hat{\bm k}_{\perp,2} \big), \label{multi}
\end{align}
where $|{\bm \alpha}_{\bm k}|^2 + |{\bm \beta}_{\bm k}|^2 \equiv w^2_{\bm k}$ and ${\bm k}=k\hat{\bm k}$. The 
three unit vectors $\hat{\bm k},\hat{\bm k}_{\perp,1}$ and $\hat{\bm k}_{\perp,2}$ form the right-handed coordinate system, 
$\hat{\bm k}_{\perp,1} \times \hat{\bm k}_{\perp,2}=\hat{\bm k}$. 

To satisfy $|{\bm n}({\bm x})|=1$ for {\it every} ${\bm x}$, 
the right-hand side of Eq.~(\ref{multi}) must have only one momentum component. 
Suppose that it has two momentum components, ${\bm k}$ and ${\bm k}^{\prime}$, 
\begin{align}
{\bm n}({\bm x}) &= \frac{w}{\sqrt{2}} \big(\cos({\bm k}{\bm x}) \hat{\bm k}_{\perp,1} - 
\sin({\bm k}{\bm x}) \hat{\bm k}_{\perp,2}\big) \nonumber \\
& \ \ + \frac{w^{\prime}}{\sqrt{2}} \big(\cos({\bm k}^{\prime}{\bm x}) \hat{\bm k}^{\prime}_{\perp,1} - 
\sin({\bm k}^{\prime}{\bm x}) \hat{\bm k}^{\prime}_{\perp,2}\big).  
\end{align}   
Without loss of generality, we can take from Eq.~(\ref{multi}) as follows,
\begin{align}
\left\{\begin{array}{c} 
{\bm k} = \frac{D}{2\lambda}{\bm e}_x \\
\hat{\bm k}_{\perp,1} = {\bm e}_y \\
\hat{\bm k}_{\perp,2} = {\bm e}_z \\
\end{array} \right. , \  
\left\{\begin{array}{l} 
{\bm k}^{\prime} = \frac{D}{2\lambda}\big(c_\gamma {\bm e}_x + s_\gamma {\bm e}_y\big) \\
\hat{\bm k}^{\prime}_{\perp,1} = c_\nu (-s_\gamma {\bm e}_x + c_\gamma {\bm e}_y) 
+ s_\nu {\bm e}_z \\
\hat{\bm k}^{\prime}_{\perp,2} = s_\nu (s_\gamma {\bm e}_x - c_\gamma {\bm e}_y) 
+ c_\nu {\bm e}_z \\
\end{array} \right. .
\end{align}
Here $c_{\nu}\equiv \cos\nu$, $s_{\nu}\equiv \sin\nu$, $c_\gamma\equiv \cos\gamma$, and $s_{\gamma}\equiv \sin\gamma$. 
Then, we have 
\begin{align}
&|{\bm n}({\bm x})|^2 = \frac{1}{2} \big(w^2+{w^\prime}^2\big)  + ww^{\prime} \Big\{ \nonumber \\
&c_{\nu}(c_\gamma-1)\cos\big(({\bm k}+{\bm k}^{\prime}){\bm x}\big) +
 c_{\nu}(c_\gamma+1)\cos\big(({\bm k}-{\bm k}^{\prime}){\bm x}\big) + \nonumber \\
& s_{\nu}(c_\gamma-1)\sin\big(({\bm k}+{\bm k}^{\prime}){\bm x}\big) 
 -  s_{\nu}(c_\gamma+1)\sin\big(({\bm k}-{\bm k}^{\prime}){\bm x}\big) \Big\}. \nonumber 
\end{align}
To make the right-hand side to be independent of ${\bm x}$, we must have 
\begin{align}
c_{\nu}(c_{\gamma}-1)= c_{\nu}(c_{\gamma}+1) = s_{\nu}(c_{\gamma}-1)= s_{\nu}(c_{\gamma}+1)=0. \label{imp}
\end{align}
Nonetheless, Eq.~(\ref{imp}) cannot be achieved by any $\gamma\in [0,2\pi)$ and $\nu \in [0,2\pi)$; this 
requires $ww^{\prime}=0$; the right hand side of Eq.~(\ref{multi}) must have only one momentum component ${\bm k}$ 
with $w_{\bm k}=\sqrt{2}$.

$C_1[{\bm n}^{\prime}]$, $C_1[{\bm n}^{\prime\prime}]$ 
and $C_{2}[{\bm n}^{\prime},{\bm n}^{\prime\prime}]$ are simultaneously maximized by  
\begin{align}
{\bm n}^{\prime}({\bm x}) ={\bm n}^{\prime\prime}({\bm x}) 
= \cos ({\bm k}{\bm x}) \hat{\bm k}_{\perp,1} - \sin({\bm k}{\bm x}) \hat{\bm k}_{\perp,2}.  \label{eq1}
\end{align}
with ${\bm k}= D/(2\lambda) \hat{\bm k}$ and $\hat{\bm k}_{\perp,1}\times \hat{\bm k}_{\perp,2}=\hat{\bm k}$. 
$\hat{\bm k}$ is an arbitrary unit vector within the $xy$-plane. With Eq.~(\ref{eq1}), the whole classical energy 
is given by $A$ and $B$ as,
\begin{align}
\frac{1}{\beta V} S[A,B] &= -\big(\alpha-\frac{2}{g}\big) (A^2+B^2) \nonumber \\
              & \ \ \ - \gamma (A^2+B^2)^2 - (A^2+B^2) \frac{D^2}{4\lambda}. 
\end{align} 
This has a global minimum at 
\begin{align}
A^2+B^2 \equiv \rho^2 = \frac{1}{2|\gamma|}\Big(\alpha-\frac{2}{g} + \frac{D^2}{4\lambda}\Big). \label{eq2}
\end{align}

To conclude, Eqs.~(\ref{eq1},\ref{eq2},\ref{rotate}) give a helicoidal order of the spin-1 exciton field as the classical 
solution at $H=0$;
\begin{align}
{\bm \phi}_{c} &= \rho e^{i\theta} \big\{{\bm e}_x \cos \omega \cos({\bm k}{\bm x}-\nu) \nonumber \\
&+ {\bm e}_y \sin \omega \cos({\bm k}{\bm x}-\nu) - {\bm e}_z \sin({\bm k}{\bm x}-\nu) \big\},  \nonumber 
\end{align} 
with 
\begin{align}
\left\{\begin{array}{c} 
{\bm k}=\frac{D}{2\lambda} (\cos \omega {\bm e}_x + \sin\omega {\bm e}_y) \\
\rho = \sqrt{\frac{1}{2|\gamma|}\Big(\alpha-\frac{2}{g} + \frac{D^2}{4\lambda}\Big)} \\ 
\end{array} \right. .  
\end{align}
The U(1) phase $\theta$, $\omega$ and $\nu$ are arbitrary. The phase $\theta$ represents a relative U(1) phase  
between the two U(1) gauges of the electron and hole. 

\subsection{$H\ne 0$ and $D \ne 0$}
The in-plane field linearly couples with the vector chirality between the real and imaginary parts of the spin-1 exciton field. 
Thereby, the classical solution at finite $h$ manifests a combined symmetry of the relative U(1) phase and the 
spin rotation, Eq.~(\ref{hs-1-c}). To see this clearly, 
let us again begin with Eqs.~(\ref{asp1},\ref{asp2}). They lead to the following  
functional for the action at finite $h$;
\begin{align}
&\frac{1}{\beta V} S[A,B,{\bm n}^{\prime}({\bm x}),{\bm n}^{\prime\prime}({\bm x})]  \nonumber \\
& \ \ \ = - \big(\alpha -\frac{2}{g} \big) \big(A^2+B^2 \big) - \gamma \big( A^4 + B^4 + 6 A^2 B^2 \big)  \nonumber \\
& \ \ -\frac{A^2}{V} C_{1}[{\bm n}^{\prime}] - \frac{B^2}{V} C_{1}[{\bm n}^{\prime\prime}] 
- \frac{1}{V} C_{3} [A,B,{\bm n}^{\prime},{\bm n}^{\prime\prime}], \nonumber   
\end{align} 
where 
\begin{align}
C_1[{\bm n}] = \int d{\bm x} & \Big[ - \lambda |{\nabla} {\bm n}^{\prime}|^2 
+ D {\bm n}^{\prime}\cdot (\nabla \times {\bm n}^{\prime}) \Big], \label{c1rr} \\
C_{3}[A,B,{\bm n}^{\prime},{\bm n}^{\prime\prime}] 
&= -4\gamma A^2B^2 \int d{\bm x} \!\ \big({\bm n}^{\prime}\cdot {\bm n}^{\prime\prime}\big)^2  \nonumber \\
& \ \ \ + 2 AB h \int d{\bm x} \!\ {\bm e}_{H}\cdot \big({\bm n}^{\prime}\times {\bm n}^{\prime\prime}\big). \label{c3r}
\end{align}
Note that for the convenience, we used the rotated frame as in Eq.~(\ref{rotate}). Thus $h{\bm e}_H$ in 
Eq.~(\ref{c3r}) is nothing but the $\pi/2$-rotation of the in-plane field around the $z$-axis. 

Let us first maximize $C_{1}[{\bm n}^{\prime}]$, $C_1[{\bm n}^{\prime\prime}]$ and 
$C_{3}[\cdots]$ with respect to ${\bm n}^{\prime}$ and 
${\bm n}^{\prime\prime}$ for fixed $A$ and $B$, and then minimize the whole action with respect to 
$A$, $B$, ${\bm n}^{\prime}$ and ${\bm n}^{\prime\prime}$. As shown above, $C_1[{\bm n}^{\prime}]$ 
and $C_1[{\bm n}^{\prime\prime}]$ are maximized by the helical orders in the rotated frame;
\begin{align}
\left\{\begin{array}{l}
{\bm n}^{\prime}({\bm x}) = \cos({\bm k}^{\prime}{\bm x}) 
\hat{\bm \alpha}^{\prime} - \sin({\bm k}^{\prime}{\bm x}) \hat{\bm \beta}^{\prime} \\
{\bm k}^{\prime} = \frac{D}{2\lambda} {\bm e}_y, \!\ \!\ 
\hat{\bm \alpha}^{\prime} = -{\bm e}_x, \!\ \!\ \hat{\bm \beta}^{\prime} = {\bm e}_z \\
\end{array} \right. , \label{n1}
\end{align}
and 
\begin{align}
\left\{\begin{array}{l}
{\bm n}^{\prime\prime}({\bm x}) = \cos({\bm k}^{\prime\prime}{\bm x}) 
\hat{\bm \alpha}^{\prime\prime} - \sin({\bm k}^{\prime\prime}{\bm x}) \hat{\bm \beta}^{\prime\prime} \\
{\bm k}^{\prime\prime} = \frac{D}{2\lambda} (\cos\omega \!\ {\bm e}_y - \sin\omega \!\ {\bm e}_x)  \\
\hat{\bm \alpha}^{\prime\prime} = \cos\nu ( - \sin\omega \!\ {\bm e}_y - \cos\omega \!\ {\bm e}_x) + \sin\nu {\bm e}_z \\
\hat{\bm \beta}^{\prime\prime} = - \sin\nu ( - \sin\omega \!\ {\bm e}_y - \cos\omega \!\ {\bm e}_x) + \cos\nu {\bm e}_z \\
\end{array} \right. . \label{n2}
\end{align}
These two give  
\begin{align}
{\bm n}^{\prime}\cdot {\bm n}^{\prime\prime} &= \frac{1}{2}(\cos \omega -1) 
\cos (({\bm k}^{\prime}+{\bm k}^{\prime\prime}){\bm x}-\nu) \nonumber \\
& \hspace{-0.2cm} + \frac{1}{2}(\cos \omega + 1)  \cos (({\bm k}^{\prime}-{\bm k}^{\prime\prime}){\bm x}+\nu), \nonumber \\
 {\bm n}^{\prime} \times {\bm n}^{\prime\prime} 
& = -\frac{1}{2}(1-\cos\omega) \sin\big(({\bm k}^{\prime}+{\bm k}^{\prime\prime}){\bm x}-\nu\big) {\bm e}_y \nonumber \\
& \hspace{-0.4cm} 
+ \frac{1}{2}(1+\cos\omega) \sin\big(({\bm k}^{\prime}-{\bm k}^{\prime\prime}){\bm x}+\nu\big) {\bm e}_y \nonumber \\
&\hspace{-1.2cm} - \frac{\sin\omega}{2}\Big[\sin\big(({\bm k}^{\prime}+{\bm k}^{\prime\prime}){\bm x}-\nu\big) 
+ \sin \big(({\bm k}^{\prime}-{\bm k}^{\prime\prime}){\bm x}+\nu\big) \Big] {\bm e}_x \nonumber \\
&  +\cdots  
\end{align}
where $\cdots$ denotes the out-of-plane component (${\bm e}_z$). Noting that ${\bm k}^{\prime\prime}=\pm {\bm k}^{\prime}$ 
for $\cos\omega=\pm$, we have
\begin{align}
\frac{1}{V} \int d{\bm x} \!\ ({\bm n}^{\prime}\cdot{\bm n}^{\prime\prime})^2 
= \left\{\begin{array}{cc} 
\frac{1}{4} (\cos^2\omega + 1) & (\cos \omega \ne \pm 1) \\
\cos^2 \nu & (\cos \omega = \pm 1) \\
\end{array}\right. ,
\end{align}
\begin{align}
\frac{1}{V} \int d{\bm x} \!\ ({\bm n}^{\prime}\times {\bm n}^{\prime\prime})_{\perp} 
= \left\{\begin{array}{cc} 
0 & (\cos \omega \ne \pm 1) \\
\sin \nu \!\ {\bm e}_y & (\cos \omega =\pm 1) \\
\end{array}\right. 
\end{align} 
Thus, without loss of generality, we can take ${\bm e}_H$ in Eq.~(\ref{c3r}) along the $+y$-direction, to 
fully maximize $C_3[{\bm n}^{\prime},{\bm n}^{\prime\prime}]$;
\begin{align}
&\frac{1}{V} C_{3}[A,B,{\bm n}^{\prime},{\bm n}^{\prime\prime}] \nonumber \\
& \hspace{0.5cm} =  \left\{\begin{array}{cc} 
|\gamma| A^2 B^2 (\cos^2 \omega + 1) & (\cos \omega \ne \pm 1) \\
4|\gamma| A^2 B^2 \cos^2\nu + 2h AB \sin\nu & (\cos \omega =\pm 1) \\
\end{array}\right. . \label{c3s}
\end{align}
Note that Eq.~(\ref{c3s}) with $\cos\omega =\pm 1$ fully maximizes $C_3[\cdots]$ for any given 
$A$ and $B$. Since Eqs.~(\ref{n1},\ref{n2}) with 
${\bm k}^{\prime}=-{\bm k}^{\prime\prime}$ ($\cos\omega=-1$) is equivalent to Eqs.~(\ref{n1},\ref{n2}) with 
${\bm k}^{\prime}={\bm k}^{\prime\prime}$ ($\cos\omega=1$) under $\nu \rightarrow \pi -\nu$, we have only to 
consider the case with ${\bm k}^{\prime}={\bm k}^{\prime\prime}$. 

\begin{figure}[t]
	\centering
	\includegraphics[width=0.7\linewidth]{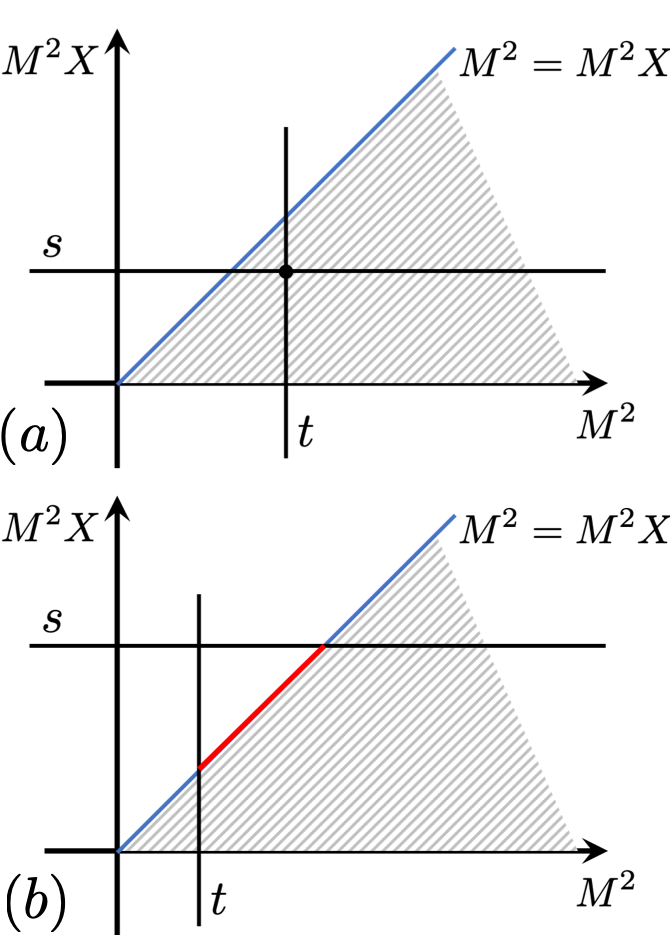}
	\caption{(color online) Locations of classical energy minima of Eq.~(\ref{action2}) in the two-dimensional 
parameter space subtended by $M^2$ and $M^2x$ $(<M^2)$. (a) $h<h_c$ (b) $h>h_c$. } 
	\label{fig:4} 
\end{figure}

When ${\bm k}^{\prime}={\bm k}^{\prime\prime}$ ($\omega=0$) in Eqs.~(\ref{n1},\ref{n2}), the total classical energy can be further 
minimized with respect to $A$, $B$ and $\nu$, the angle between ${\bm \alpha}^{\prime}$ and ${\bm \alpha}^{\prime\prime}$;
\begin{align}
\frac{1}{\beta V} S[A,B,\nu] &= - \Big(\alpha-\frac{2}{g}+\frac{D^2}{4\lambda}\Big) (A^2+B^2) \nonumber \\
&  \hspace{-1.7cm} - \gamma (A^2+B^2)^2 + 4|\gamma| A^2 B^2 \sin^2\nu - 2h A B \sin \nu. \nonumber 
\end{align} 
Namely, take $(A,B) \equiv M(\cos\theta,\sin\theta)$, and minimize the energy with respect to $M$ and 
$x\equiv \sin2\theta\sin\nu$ $(\le 1)$, 
\begin{align}
\frac{S}{\beta V} &= |\gamma| \big[ (M^2 -t)^2 + (M^2 x - s)^2 - (t^2+s^2)\big]. \label{action2}
\end{align} 
Here $s$ and $t$ are given by
\begin{align}
t \equiv \frac{h_c}{2|\gamma|}, \ \ 
s \equiv   \frac{h}{2|\gamma|}, \ \ h_c \equiv \alpha-\frac{2}{g}+\frac{D^2}{4\lambda}. 
\end{align}
Since $M^2 x\le M^2$, the energy has two different minima, depending on whether $s<t$ ($h<h_c$) or 
$s>t$ ($h>h_c$). When $h\le h_c$ (Fig.~\ref{fig:4}(a)), the energy has a minimum at 
\begin{eqnarray}
M^2 = t = \frac{h_c}{2|\gamma|}, \!\ M^2 x =s= \frac{h}{2|\gamma|}. \label{solution1}
\end{eqnarray}
When $h\ge h_c$ (Fig.~\ref{fig:4}(b)), the energy must  be minimized along $x=1$. Substituting 
$x=1$ into Eq.~(\ref{action2}), one can see that it has a minimum at  
\begin{align}
M^2 = \frac{1}{2} (t+s)= \frac{h_c+h}{4|\gamma|}.  \label{solution2}
\end{align}

In conclusion, the classical ground state configuration in the presence of the finite in-plane field is characterized by 
two helicoid orders of ${\bm \phi}^{\prime}_c({\bm x})$ and ${\bm \phi}^{\prime\prime}_c({\bm x})$. 
When the in-plane field is along the $+x$-direction, they take the following forms.  For $h<h_c$, 
\begin{align}
\left\{\begin{array}{c}
{\bm \phi}^{\prime}_c({\bm x}) = \rho \cos\theta \big(\cos(Ky){\bm e}_y -\sin(Ky) {\bm e}_z\big) \\
{\bm \phi}^{\prime\prime}_c({\bm x}) = \rho \sin\theta \big(\cos(Ky-\nu){\bm e}_y -
\sin(Ky-\mu) {\bm e}_z\big) \\ 
K = \frac{D}{2\lambda}, \!\ \!\ \rho =  \sqrt{\frac{h_c}{2|\gamma|}}, \!\ \!\ 
\sin\nu\sin2\theta=\frac{h}{h_c} \\
\end{array}\right.  . \label{hlhc}
\end{align} 
For $h>h_c$, 
\begin{align}
\left\{\begin{array}{c}
{\bm \phi}^{\prime}_c({\b x})+i{\bm \phi}^{\prime\prime}_c({\bm x})
= e^{iKy} \frac{\rho^{\prime}}{\sqrt{2}} ({\bm e}_y+i{\bm e}_z)  \\
K = \frac{D}{2\lambda}, \!\ \!\ \rho^{\prime}=\sqrt{\frac{h_c+h}{4|\gamma|}} \\
\end{array}\right. . \label{hghc} 
\end{align} 

\section{derivation of linearized EOM for fluctuation of the spin-1 exciton field}
In this appendix, we derive a linearized equation of motion (EOM) for a fluctuation of 
the spin-1 exciton field around the helicoidal structure (Eqs.~(\ref{hlhc},\ref{hghc})). 
We first take a functional derivative of the effective action (Eq.~(\ref{action})), 
to derive a coupled nonlinear EOMs for real and imaginary parts of the spin-1
exciton field; 
\begin{widetext}
\begin{align}
|\eta| \partial_{t} \phi^{\prime\prime}_x &- \Big(\alpha-\frac{2}{g}\Big) \phi^{\prime}_x 
+ 2|\gamma| |{\bm \phi}^{\prime}|^2 \phi^{\prime}_x + 6|\gamma|  |{\bm \phi}^{\prime\prime}|^2 
\phi^{\prime}_x - 4|\gamma| ({\bm \phi}^{\prime}\cdot {\bm \phi}^{\prime\prime}) \phi^{\prime\prime}_x 
+ D\partial_x \phi^{\prime}_z - \lambda {\bm \nabla}^2 \phi^{\prime}_x = 0, \nonumber \\
-|\eta| \partial_{t} \phi^{\prime}_x &- \Big(\alpha-\frac{2}{g}\Big) \phi^{\prime\prime}_x 
+ 2|\gamma| |{\bm \phi}^{\prime\prime}|^2 \phi^{\prime\prime}_x + 6|\gamma|  |{\bm \phi}^{\prime}|^2 
\phi^{\prime\prime}_x - 4|\gamma| ({\bm \phi}^{\prime}\cdot{\bm \phi}^{\prime\prime}) \phi^{\prime}_x 
+ D\partial_x \phi^{\prime\prime}_z - \lambda {\bm \nabla}^2 \phi^{\prime\prime}_x = 0, \nonumber \\
|\eta| \partial_{t} \phi^{\prime\prime}_y &- \Big(\alpha-\frac{2}{g}\Big) \phi^{\prime}_y 
+ 2|\gamma| |{\bm \phi}^{\prime}|^2 \phi^{\prime}_y + 6|\gamma|  |{\bm \phi}^{\prime\prime}|^2 
\phi^{\prime}_y - 4|\gamma| ({\bm \phi}^{\prime}\cdot {\bm \phi}^{\prime\prime}) \phi^{\prime\prime}_y 
+ D\partial_y \phi^{\prime}_z - \lambda {\bm \nabla}^2 \phi^{\prime}_y -h \phi^{\prime\prime}_z = 0, \nonumber \\
-|\eta| \partial_{t} \phi^{\prime}_y &- \Big(\alpha-\frac{2}{g}\Big) \phi^{\prime\prime}_y 
+ 2|\gamma| |{\bm \phi}^{\prime\prime}|^2 \phi^{\prime\prime}_y + 6|\gamma|  |{\bm \phi}^{\prime}|^2 
\phi^{\prime\prime}_y - 4|\gamma| ({\bm \phi}^{\prime}\cdot{\bm \phi}^{\prime\prime}) \phi^{\prime}_y 
+ D\partial_y \phi^{\prime\prime}_z - \lambda {\bm \nabla}^2 \phi^{\prime\prime}_y + h\phi^{\prime}_z = 0, \nonumber \\ 
|\eta| \partial_{t} \phi^{\prime\prime}_z &- \Big(\alpha-\frac{2}{g}\Big) \phi^{\prime}_z 
+ 2|\gamma| |{\bm \phi}^{\prime}|^2 \phi^{\prime}_z + 6|\gamma|  |{\bm \phi}^{\prime\prime}|^2 
\phi^{\prime}_z - 4|\gamma| ({\bm \phi}^{\prime}\cdot {\bm \phi}^{\prime\prime}) \phi^{\prime\prime}_z 
- D{\bm  \nabla} \cdot {\bm \phi}^{\prime} - \lambda {\bm \nabla}^2 \phi^{\prime}_z  + h \phi^{\prime\prime}_y = 0, \nonumber \\
-|\eta| \partial_{t} \phi^{\prime}_z &- \Big(\alpha-\frac{2}{g}\Big) \phi^{\prime\prime}_z 
+ 2|\gamma| |{\bm \phi}^{\prime\prime}|^2 \phi^{\prime\prime}_z + 6|\gamma|  |{\bm \phi}^{\prime}|^2 
\phi^{\prime\prime}_z - 4|\gamma| ({\bm \phi}^{\prime}\cdot{\bm \phi}^{\prime\prime}) \phi^{\prime}_z 
- D {\bm  \nabla} \cdot {\bm \phi}^{\prime\prime} - \lambda {\bm \nabla}^2 \phi^{\prime\prime}_z - h\phi^{\prime}_y = 0, \nonumber 
\end{align}
\end{widetext}
with ${\bm \nabla}\equiv (\partial_x,\partial_y,0)$. 
Here we have replaced the imaginary time $\tau$ by the real time $t$; $i\partial_{\tau}=\partial_t$. 
The classical configurations given by Eqs.~(\ref{hlhc},\ref{hghc}) are static solutions 
of the nonlinear EOMs. We thus introduce a small fluctuation of the exciton field around the classical configuration, 
${\bm \phi}^{\prime}({\bm x}) \equiv {\bm \phi}^{\prime}_{c}({\bm x})+\delta{\bm \phi}^{\prime}({\bm x})$ 
and ${\bm \phi}^{\prime\prime}({\bm x}) \equiv {\bm \phi}^{\prime\prime}_{c}({\bm x})+\delta{\bm \phi}^{\prime\prime}({\bm x})$ 
and linearize the EOMs with respect to the fluctuations, $\delta {\bm \phi}^{\prime}$ and $\delta {\bm \phi}^{\prime\prime}$.
For $h\le h_c$, the linearized coupled EOMs for ${\bm \phi}^{\dagger} \equiv 
\delta {\bm \phi}^{\prime}-i\delta {\bm \phi}^{\prime\prime}$ and 
${\bm \phi} \equiv \delta {\bm \phi}^{\prime}+i\delta {\bm \phi}^{\prime\prime}$ are given by
\begin{widetext}
\begin{align}
&i|\eta|\partial_{t} \phi_x + \alpha_{\bm \nabla} \phi_x + 2|\gamma|\rho^2 F_0 
\phi^{\dagger}_x - D\partial_x \phi_z = 0, \nonumber \\
&i|\eta|\partial_{t} \phi^{\dagger}_x - \alpha_{\bm \nabla} \phi^{\dagger}_x 
- 2|\gamma|\rho^2 F^{*}_0 \phi_x + D\partial_x \phi^{\dagger}_z = 0, \nonumber \\
&i|\eta|\partial_{t} \phi_y + \alpha_{\bm \nabla} \phi_y  - (D\partial_y-ih) \phi_{z}
- C_y \phi^{\dagger}_y + S_y \phi^{\dagger}_z = 0, \nonumber \\
&i|\eta|\partial_{t} \phi^{\dagger}_y - \alpha_{\bm \nabla} \phi^{\dagger}_y  + (D\partial_y+ih) \phi^{\dagger}_{z}
+ C^{*}_y \phi_y - S^{*}_y \phi_z = 0, \nonumber \\
&i|\eta|\partial_{t} \phi_z + \alpha_{\bm \nabla} \phi_z  + D{\bm \nabla} {\bm \phi} -ih \phi_{y}
+ C_y \phi^{\dagger}_z + S_y \phi^{\dagger}_y = 0, \nonumber \\
&i|\eta|\partial_{t} \phi^{\dagger}_z - \alpha_{\bm \nabla} \phi^{\dagger}_z  - D {\bm \nabla} {\bm \phi}^{\dagger}  
- ih \phi^{\dagger}_y - C^{*}_y \phi_z - S^{*}_y \phi_y = 0. \nonumber 
\end{align}
where $\rho$, $\alpha_{\bm \Delta}$, $F_0$, $C_y$, and $S_y$ are defined in Eqs.~(\ref{hs-0-c},\ref{a1},\ref{f0},\ref{cy},\ref{sy}) 
respectively. For $h\ge h_c$, the linearized EOMs are given by
\begin{align}
&i|\eta|\partial_{t} \phi_x + \alpha^{\prime}_{\bm \nabla} \phi_x - D\partial_x \phi_z = 0, \nonumber \\
&i|\eta|\partial_{t} \phi^{\dagger}_x - \alpha^{\prime}_{\bm \nabla} \phi^{\dagger}_x 
+ D\partial_x \phi^{\dagger}_z = 0, \nonumber \\
&i|\eta|\partial_{t} \phi_y + \alpha^{\prime}_{\bm \nabla} \phi_y  - (D\partial_y+ih-i2\zeta) \phi_{z}
- \zeta e^{2iKy} (\phi^{\dagger}_y +i  \phi^{\dagger}_z) = 0, \nonumber \\
&i|\eta|\partial_{t} \phi^{\dagger}_y - \alpha^{\prime}_{\bm \nabla} \phi^{\dagger}_y  + (D\partial_y-ih+i2\zeta) 
\phi^{\dagger}_{z} + \zeta e^{-2iKy} (\phi_y - i \phi_z) = 0, \nonumber \\
&i|\eta|\partial_{t} \phi_z + \alpha^{\prime}_{\bm \nabla} \phi_z  + D{\bm \nabla} {\bm \phi} + (ih-2i\zeta) \phi_{y}
+ \zeta e^{2iKy} (\phi^{\dagger}_z -i \phi^{\dagger}_y) = 0, \nonumber \\
&i|\eta|\partial_{t} \phi^{\dagger}_z - \alpha^{\prime}_{\bm \nabla} \phi^{\dagger}_z  - D {\bm \nabla} {\bm \phi}^{\dagger}  
+ (ih-2i\zeta) \phi^{\dagger}_y - \zeta e^{-2iKy}
 (\phi_z + i  \phi_y) = 0. \nonumber 
\end{align}
\end{widetext}
where $\rho^{\prime}$, $\alpha^{\prime}_{\bm \nabla}$, and $\zeta$ are defined in Eqs.~(\ref{hs-2-b},\ref{a2},\ref{zeta}). 

As indicated by the form of the effective action, 
${\bm \phi}^{\dagger}$ and ${\bm \phi}$ play role of spin-1 
boson creation and annihilation operator respectively. Therefore, 
the linearized EOMs should take a form of generalized eigenvalue equation with 
a bosonic BdG Hamiltonian;
\begin{align}
|\eta| i\partial_{t} \left(\begin{array}{c}
\phi_x  ({\bm x})\\
\phi_y  ({\bm x})\\
\phi_z  ({\bm x})\\
\phi^{\dagger}_x ({\bm x}) \\
\phi^{\dagger}_y ({\bm x}) \\
\phi^{\dagger}_z  ({\bm x})\\
\end{array}\right) = {\bm \tau}_3 \hat{\bm H}_{\rm BdG}({\bm \nabla},{\bm x}) \!\ \left(\begin{array}{c}
\phi_x ({\bm x})\\
\phi_y ({\bm x}) \\
\phi_z  ({\bm x})\\
\phi^{\dagger}_x  ({\bm x})\\
\phi^{\dagger}_y ({\bm x}) \\
\phi^{\dagger}_z  ({\bm x})\\
\end{array}\right), 
\end{align} 
where $\hat{\bm H}_{\rm BdG}$ is an Hermitian operator
\begin{align}
\hat{\bm H}^{\dagger}_{\rm BdG}(\partial_x,\partial_y,x,y) = \hat{\bm H}_{\rm BdG}(-\partial_x,-\partial_y,x,y). 
\label{hermite}
\end{align}
In fact, Eq.~(\ref{hermite}) holds true for eqs.~(\ref{bdg1},\ref{bdg2}).  

\section{evaluation of local magnetic moment and local charge density in the electron layer}
In this appendix, we calculate local magnetic moment and local charge density in the 
electron layer, that are induced by the helicoidal excitonic order ($h<h_c$). To this end,  
let us begin with the following temperature Green's function~\cite{fetter03},  
\begin{align}
G^a_{\alpha\beta}({\bm x},\tau;{\bm x}^{\prime},\tau^{\prime}) 
= - \frac{{\rm Tr}\big[e^{-\beta K} {\cal T}_{\tau} \big\{a_{\alpha}({\bm x},\tau) 
a^{\dagger}_{\beta}({\bm x}^{\prime},\tau^{\prime})\big\}\big]}{{\rm Tr}[e^{-\beta K}]}
\end{align}  
where 
\begin{align}
a_{\alpha}({\bm x},\tau) &\equiv e^{K\tau} a_{\alpha}({\bm x}) e^{-K\tau}, \nonumber \\
a^{\dagger}_{\alpha}({\bm x},\tau) &\equiv e^{K\tau} a^{\dagger}_{\alpha}({\bm x}) e^{-K\tau},  \nonumber 
\end{align}
with $K \equiv H_{0}-\mu N + H^{\prime} \equiv K_0 + H^{\prime}$ and 
\begin{align}
- H^{\prime} &= \int d{\bm x} \!\ \big({\bm \phi}^{\prime}_c({\bm x})-i{\bm \phi}^{\prime\prime}_c({\bm x})\big) 
{\bm b}^{\dagger}({\bm x}) {\bm \sigma} {\bm a}({\bm x}) \nonumber \\ 
& \ \  + \int d{\bm x} \!\ \big({\bm \phi}^{\prime}_c({\bm x})+i{\bm \phi}^{\prime\prime}_c({\bm x})\big) 
{\bm a}^{\dagger}({\bm x}) {\bm \sigma} {\bm b}({\bm x}).  \nonumber 
\end{align}
The classical configuration of the spin-1 exciton field for $h<h_c$ 
is given by Eqs.~(\ref{hs-1-a},\ref{hs-1-b}). In the 
momentum space, $K_0$ and $H^{\prime}$ are 
given by
\begin{align}
K_0 &= \sum_{\bm k} a^{\dagger}_{\bm k} ( \xi^{a}_{\bm k} {\bm \sigma}_0 + \xi_e (k_y{\bm \sigma}_x 
- k_x {\bm \sigma}_y) + H {\bm \sigma}_x ) a_{\bm k} \nonumber \\
& \ + \sum_{\bm k} b^{\dagger}_{\bm k} ( \xi^{b}_{\bm k} {\bm \sigma}_0 + \Delta_h (k_x{\bm \sigma}_x 
+ k_y {\bm \sigma}_y) + H {\bm \sigma}_x ) b_{\bm k} \nonumber \\
H^{\prime} & = - \frac{\rho}{2} \sum_{{\bm k}_1,{\bm k}_2}\!\nonumber \\ 
& \hspace{-0.5cm}   \Big\{ 
{\bm b}^{\dagger}_{{\bm k}_1} ( {\bm \sigma}_y +i{\bm \sigma}_z) {\bm a}_{{\bm k}_2} \delta_{{\bm k}_1,{\bm k}_2+K{\bm e}_y} 
\big(\cos\theta - ie^{-i\nu}\sin\theta\big) \nonumber \\ 
& \hspace{-0.2cm} + {\bm b}^{\dagger}_{{\bm k}_1} ( {\bm \sigma}_y -i {\bm \sigma}_z) 
{\bm a}_{{\bm k}_2} \delta_{{\bm k}_1,{\bm k}_2-K{\bm e}_y} 
\big(\cos\theta - ie^{i\nu}\sin\theta\big) \nonumber \\ 
& \hspace{-0.4cm} +  {\bm a}^{\dagger}_{{\bm k}_1} ( {\bm \sigma}_y +i {\bm \sigma}_z) 
{\bm b}_{{\bm k}_2} \delta_{{\bm k}_1,{\bm k}_2+K{\bm e}_y} 
\big(\cos\theta + ie^{-i\nu}\sin\theta\big) \nonumber \\ 
&\hspace{-0.6cm} 
+ {\bm a}^{\dagger}_{{\bm k}_1} ({\bm \sigma}_y -i {\bm \sigma}_z) 
{\bm b}_{{\bm k}_2} \delta_{{\bm k}_1,{\bm k}_2-K{\bm e}_y} 
\big(\cos\theta + ie^{i\nu}\sin\theta\big) \Big\}, \nonumber 
\end{align}
with $\xi^{a/b}_{\bm k} \equiv {\cal E}_{a/b}({\bm k})-\mu$. 

Using the standard Feynman-Dyson perturbation theory~\cite{fetter03}, 
we evaluate the temperature Green function up to the lowest order 
in $H^{\prime}$. Since $H^{\prime}$ connects between electron and hole but it does not between electron and 
electron or between hole and hole, the lowest order starts from the second order in $H^{\prime}$;
\begin{align}
G^{a}_{\alpha\beta}({\bm x},\tau:{\bm x}^{\prime},\tau^{\prime}) 
&=   \nonumber  \\
&\hspace{-1.8cm} \frac{1}{V} \sum_{{\bm q},{\bm q}^{\prime}} 
\frac{1}{\beta} \sum_{i\omega_n}
e^{i{\bm q}{\bm x}-i{\bm q}^{\prime}{\bm x}^{\prime}} e^{-i\omega_n(\tau-\tau^{\prime})} 
{G}^{a}_{\alpha\beta}({\bm q},{\bm q}^{\prime}:i\omega_n),  \label{fourier2} \\
G^{a}_{\alpha\beta}
({\bm q},{\bm q}^{\prime}:i\omega_n) & = \delta_{{\bm q}-2K{\bm e}_y,{\bm q}^{\prime}} 
{G}^{a}_{-}({\bm q}-K{\bm e}_y,i\omega_n) 
\nonumber \\ 
& \hspace{-2.1cm} + \delta_{{\bm q},{\bm q}^{\prime}} {G}^{a}_{0}({\bm q},i\omega_n) + 
 \delta_{{\bm q}+2K{\bm e}_y,{\bm q}^{\prime}} {G}^{a}_{+}({\bm q}+K{\bm e}_y,i\omega_n).  \label{3-cont}
\end{align}
Here 
\begin{align}
{G}^{a}_{\mp}({\bm q},i\omega_n) &= \Big(\frac{\rho}{2}\Big)^2 
(\cos^2\theta + e^{\mp i2\nu} \sin^2\theta) \!\ 
{g}^a_0({\bm q}\pm K{\bm e}_y,i\omega_n) \nonumber \\
& \hspace{-1.7cm} ({\bm \sigma}_y\pm i {\bm \sigma}_z) 
 {g}^b_0({\bm q},i\omega_n) ({\bm \sigma}_y\pm i {\bm \sigma}_z) 
{g}^a_0({\bm q}\mp K {\bm e}_y,i\omega_n), \nonumber \\
{G}^{a}_{0}({\bm q},i\omega_n) &= \Big(\frac{\rho}{2}\Big)^2 
\sum_{\sigma=\pm} (1 +\sigma \sin\nu \sin 2\theta) \!\ 
{g}^a_0({\bm q},i\omega_n) \nonumber \\
& \hspace{-1.7cm} ({\bm \sigma}_y + i\sigma {\bm \sigma}_z) 
 {g}^b_0({\bm q} -\sigma K{\bm e}_y,i\omega_n) ({\bm \sigma}_y - \sigma i {\bm \sigma}_z) 
{g}^a_0({\bm q},i\omega_n), \nonumber
\end{align}
and 
\begin{align}
{g}^{a}_0({\bm q},i\omega_n) &= \frac{(i\omega_n-\xi^a_{\bm q}){\bm \sigma}_0 
+ \xi_e (q_y {\bm \sigma}_x - q_x {\bm \sigma}_y) + H{\bm \sigma}_x}
{(i\omega_n-\xi^a_{\bm q})^2- \big[(\xi_e q_x)^2 + (\xi_e q_y + H)^2\big]}, \nonumber \\  
{g}^{b}_0({\bm q},i\omega_n) &= \frac{(i\omega_n-\xi^b_{\bm q}){\bm \sigma}_0 
+ \Delta_h (q_x {\bm \sigma}_x + q_y {\bm \sigma}_y) + H{\bm \sigma}_x}
{(i\omega_n-\xi^b_{\bm q})^2- \big[(\Delta_h q_y)^2 + (\Delta_h q_x + H)^2\big]}. \nonumber 
\end{align}

The local magnetic moment and charge density in the electron layer is calculated from the 
Green function,
\begin{align}
\left\{\begin{array}{c}
\rho^e({\bm x}) = {\rm Tr}\big[{G}^{a}({\bm x},\tau;{\bm x},\tau+0)\big]  \\
{\bm m}^{e}({\bm x}) = \frac{1}{2} {\rm Tr}\big[{\bm \sigma}{G}^{a}({\bm x},\tau;{\bm x},\tau+0)\big] \\ 
\end{array}\right. \label{sub} 
\end{align}
When Eqs.~(\ref{3-cont},\ref{fourier2}) are substituted into Eq.~(\ref{sub}), the first and third terms 
in Eq.~(\ref{3-cont}) give rise to helicoidal {\it spin} density wave in the $yz$ plane, 
while the second term in Eq.~(\ref{3-cont}) leads to uniform charge density and magnetic moment along the 
in-plane Zeeman field ($x$-direction). In the leading order in $\Delta_h$, they are given by
\begin{align}
\rho^{e}({\bm x}) &= C + {\cal O}(\Delta^2_h) \label{rhoe} \\
{\bm m}^{e}({\bm x}) &= A \big(\cos^2\theta \cos(2Ky) + \sin^2\theta \cos(2Ky-2\nu) \big) {\bm e}_y \nonumber \\
& - B \big(\cos^2\theta \sin(2Ky) + \sin^2\theta \sin(2Ky-2\nu) \big) {\bm e}_z \nonumber \\
& + \frac{1}{2} D {\bm e}_x + {\cal O}(\Delta^2_h) \label{me} 
\end{align}
where 
\begin{align}
\left(\begin{array}{c} 
A \\
B \\
\end{array}\right) &\equiv \frac{\rho^2 \Delta_h}{V} \sum_{\bm q} \frac{1}{\beta} \sum_{i\omega_n} e^{i\omega_n 0+} 
q_y f({\bm q},H) 
\left(\begin{array}{c} 
t+s \\
t-s \\
\end{array}\right),  
\end{align}
and 
\begin{align}
\left(\begin{array}{c}
C \\
D \\
\end{array}\right) & \equiv   \frac{\rho^2}{V} \sum_{\bm q} \frac{1}{\beta} \sum_{i\omega_n} e^{i\omega_n 0+} \sum_{\sigma=\pm} 
(1+\sigma \sin\nu\sin 2\theta)  \nonumber \\
&\hspace{0.5cm} 
(i\omega_n-\xi^b_{\bm q}-H) g_{\sigma}({\bm q},H) \left(\begin{array}{c} 
u_{\sigma}+s \\
u_{\sigma}-s \\
\end{array}\right), 
\end{align}
with 
\begin{widetext}
\begin{align}
f({\bm q},H) &\equiv \frac{1}{(i\omega_n-\xi^{b}_{\bm q})^2-H^2} \prod_{\sigma=\pm}
\frac{1}{(i\omega_n-\xi^a_{{\bm q}+\sigma})^2-[(\xi_e q_x)^2 + (\xi_e(q_y+\sigma K)+H)^2]}, \nonumber \\
t &\equiv [(i\omega_n-\xi^a_{{\bm q}+})+(\xi_e(q_y+K)+H)][(i\omega_n-\xi^a_{{\bm q}-})-(\xi_e(q_y-K)+H)],  \ \ 
s  \equiv (\xi_e q_x)^2, \nonumber \\
g_{\sigma}({\bm q},H) & \equiv \frac{1}{(i\omega_n-\xi^{b}_{\bm q})^2-H^2} 
\bigg(\frac{1}{(i\omega_n-\xi^a_{{\bm q}+\sigma})^2- [(\xi_e q_x)^2 + (\xi_e(q_y+\sigma K)+H)^2]}\bigg)^2, \nonumber \\
u_{\sigma}& \equiv [(i\omega_n-\xi^a_{{\bm q}+\sigma})+(\xi_e(q_y+\sigma K)+H)]^2, 
\end{align} 
\end{widetext}
and ${\bm q}+\sigma\equiv (q_x,q_y+\sigma K)$ ($\sigma=\pm$). 
Eqs.~(\ref{rhoe},\ref{me}) conclude that up to the first order in $\Delta_h$, the helicoidal excitonic order under the in-plane 
Zeeman field (along $x$) induces the uniform charge density and uniform magnetization (along $x$) as well as the helicoidal magnetic 
order within the $yz$ plane in the electron layer. A finite uniform charge density induced by the helicoidal excitonic order 
suggests that the low-energy collective modes in the excitonic phase can couple electrically with 
external electromagnetic waves. The helicoidal magnetic texture within the $yz$ plane suggests that the helicoidal 
structure can be seen by the magneto-optical Kerr spectroscopy.

\end{document}